\documentclass[10pt,twocolumn,letterpaper]{article}

\usepackage{cvpr}
\usepackage{times}
\usepackage{epsfig}
\usepackage{graphicx}
\usepackage{amsmath}
\usepackage{amssymb}
\usepackage{subcaption}
\usepackage{xcolor}

\usepackage{makecell}

\usepackage[pagebackref=true,breaklinks=true,letterpaper=true,colorlinks,bookmarks=false]{hyperref}

\cvprfinalcopy 

\def\cvprPaperID{14} 

\begin{document}

\title{Complexity Evaluation of Parallel Execution of\\
the {RAPiD} Deep-Learning Algorithm on {Intel CPU}}

\author{Dominic Konrad,\\
ECE Department\\
University of California, Los Angeles\\
{\tt\small djkonrad@g.ucla.edu}\\
\and
Zhihao Duan,\\
ECE Department\\
Purdue University\\
{\tt\small duan90@purdue.edu}\\
\and
Mertcan Cokbas, Prakash Ishwar\\
ECE Department\\
Boston University\\
{\tt\small \{mcokbas@bu.edu,pi\}@bu.edu}
}

\maketitle

\begin{abstract}
\vskip -0.3cm
Knowing how many and where are people in various indoor spaces is critical for reducing HVAC energy waste, space management, spatial analytics and in emergency scenarios. While a range of technologies have been proposed to detect and track people in large indoor spaces, ceiling-mounted fisheye cameras have recently emerged as strong contenders. Currently, RAPiD (Rotation-Aware People Detection) is the state-of-the-art algorithm for people detection in images captured by fisheye cameras. However, in large spaces several overhead fisheye cameras are needed to assure high accuracy of counting and thus multiple instances of RAPiD must be executed simultaneously.

This report evaluates inference time when multiple instances of RAPiD run in parallel on an Ubuntu NUC PC with Intel I7 8559U CPU.  We consider three mechanisms of CPU-resource allocation to handle multiple instances of RAPiD: 1) managed by Ubuntu, 2) managed by user via operating-system calls to assign logical cores, and 3) managed by user via PyTorch-library calls to limit the number of threads used by PyTorch.
Each scenario was evaluated on 300 images.
The experimental results show, that when one or two instances of RAPiD are executed in parallel all three approaches result in similar inference times of 1.8sec and 3.2sec, respectively. However, when three or more instances of RAPiD run in parallel, limiting the number of threads used by PyTorch results in the shortest inference times.
On average, RAPiD completes inference of 2 images simultaneously in about 3sec, 4 images in 6sec and 8 images in less than 14sec.
This is important for real-time system design. In HVAC-application scenarios, with a typical reaction time of 10-15min, a latency of 14sec is negligible so a single 8559U CPU can support 8 camera streams thus reducing the system cost. However, in emergency scenarios, when time is of essence, a single CPU may be needed for each camera to reduce the latency to 1.8sec. 
\end{abstract}

\section{Introduction}  

Occupancy sensing, that is understanding how many and where people are in a building, is a key technology for:
\begin{itemize}
	\item reducing HVAC energy waste (air flow matched to occupancy),
	\item space management (quantification of space usage to reduce rental costs),
	\item spatial analytics (quantification of customer flow in retail spaces),
	\item emergency scenarios (fire, chemical hazard, active shooter).
\end{itemize}

Among visual occupancy-sensing methods, standard video cameras mounted high on walls have been most common. However, in order to overcome their relatively narrow field of view (FOV), recently overhead fisheye cameras have become the sensing modality of choice for their wide FOV (typically $360^\circ\times 180^\circ$) and overhead viewpoint that reduces mutual occlusions between people, thus simplifying detection.

While numerous deep-learning (DL) algorithms have been developed for people detection using standard surveillance cameras, they do not reliably detect people in overhead fisheye images due to camera viewpoint and geometric distortions.
Since cameras are mounted above the scene and looking down, standing people are radially oriented in overhead fisheye images. On the other hand, standard people-detection algorithms, such as YOLO \cite{yolo}, R-CNN \cite{r-cnn}, etc., are designed for person detection with bounding boxes aligned to image axes, thus struggling with radial orientations. Also, fisheye cameras are equipped with a wide-angle lens that has strong geometric distortions, especially at FOV periphery, causing additional difficulties with reliable people detection.

Recently, a people-detection DL algorithm, called RAPiD (Rotation-Aware People Detection) \cite{zhihao}, has been developed to accommodate rotation of each person's bounding box. RAPiD is based on YOLOv3 \cite{yolov3,yolo}, but it could be extended to its newer versions such YOLOv5, YOLOx, etc. Currently, RAPiD is a state-of-the-art DL algorithm for finding people in RGB images captured by ceiling-mounted fisheye cameras. 

This project aims at quantifying RAPiD's computational complexity on a contemporary CPU in order to inform the design of a {\it real-time system for occupancy sensing in large indoor spaces}. Such spaces require the use of multiple overhead fisheye cameras. In addition to the cost of cameras, an important factor is the cost of computing hardware. In order to inform design decisions, this project evaluates three parallel deployment strategies of RAPiD (to support multiple cameras) on a single CPU in terms of inference latency.

\section{Problem statement}  

The considered system architecture consists of $N$ fisheye cameras mounted overhead in a large space (e.g., university auditorium, convention hall) and a modern NUC (Next Unit of Computing) PC, all connected to a private PoE (Power over Ethernet) LAN that provides both communication and power. The NUC PC collects video frames from the cameras and runs RAPiD on each frame. The main questions that this project attempts to answer are:
\begin{itemize}
	\item How long does it take for RAPiD to complete people detection for a single video frame?
	\item How does the above inference time scale with the number of RAPiD instances running simultaneously?
	\item How does the inference time scale with scene complexity (number of people)?
	\item Can hyperthreading help speed up the execution of RAPiD?
\end{itemize}

\section{Experimental setup}  

Our setup consists of the following components:
\begin{itemize}
	\item Axis M3057-PLVE fisheye cameras,
	\item NUC PC with Intel Quad-Core I7 8559U CPU (2.7-4.5 GHz), Crucial 16GB RAM (DDR4, 2,400 MHz), Samsung 970 EVO 500 MB SSD,
	\item 1 Gb/s LAN.
\end{itemize}
The NUC is a small form-factor PC (117 $\times$ 112 $\times$ 51mm) that takes little space and can be easily mounted in a suspended ceiling close to the cameras. It is powered by an external "laptop-style" 19V power supply, that supplies 28W at base clock frequency of 2.7 GHz. The NUC PC runs Linux Ubuntu 18.04.4 which is accessed remotely using SSH (no GUI).

\begin{figure}[!htb]
  \centerline{\epsfig{figure=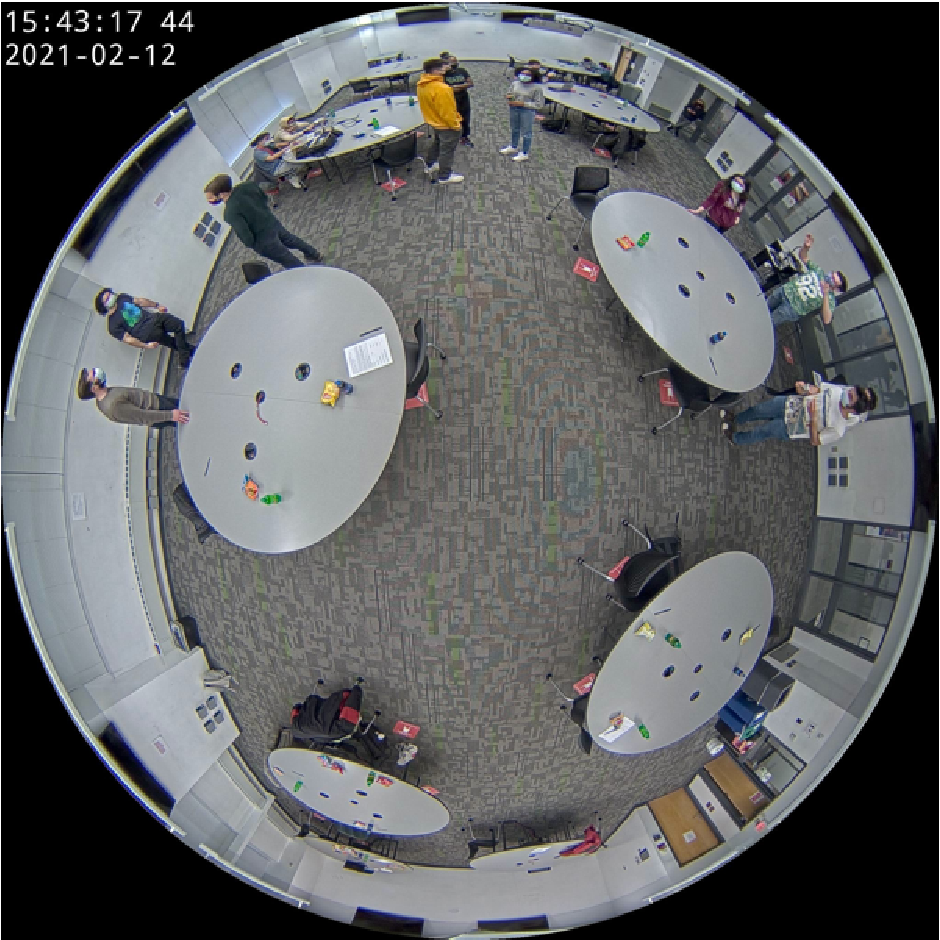,width=6cm}}
  \medskip
  \centerline{(a) Medium-complexity scenario (images 1-100): 14 people}
  \medskip
  \centerline{\epsfig{figure=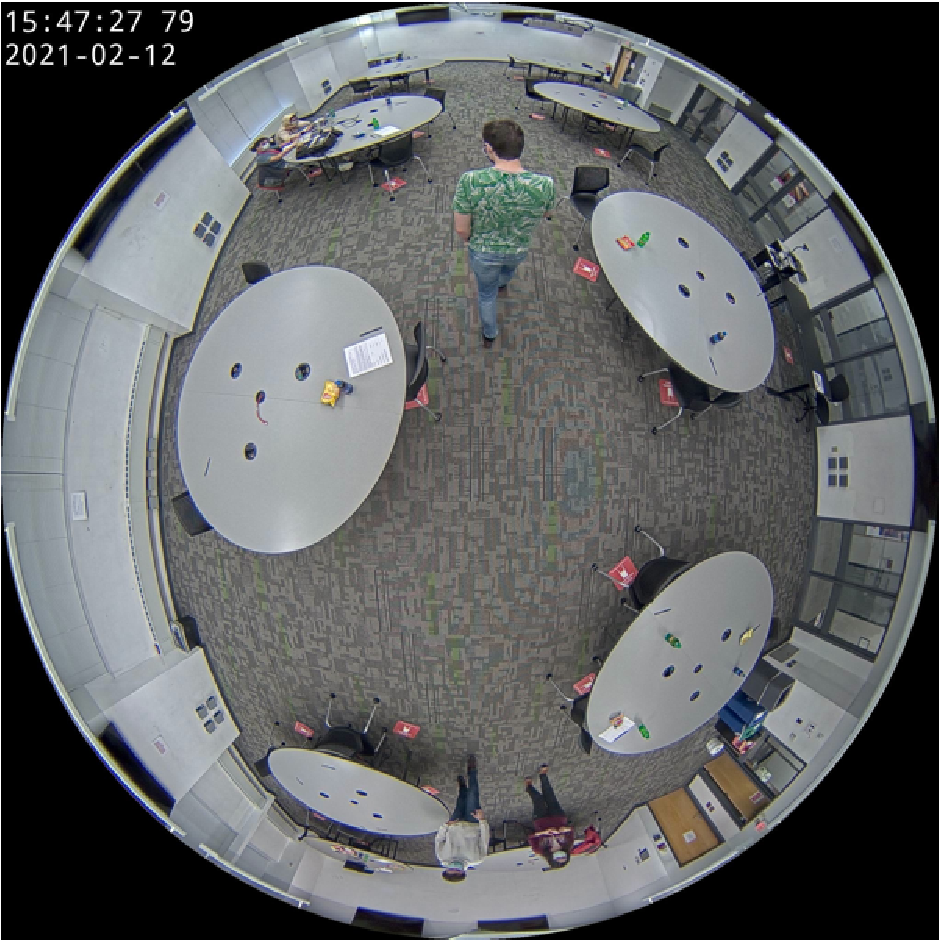,width=6cm}}
  \centerline{(b) Low-complexity scenario (images 101-200): 5-6 people}
  \medskip
  \centerline{\epsfig{figure=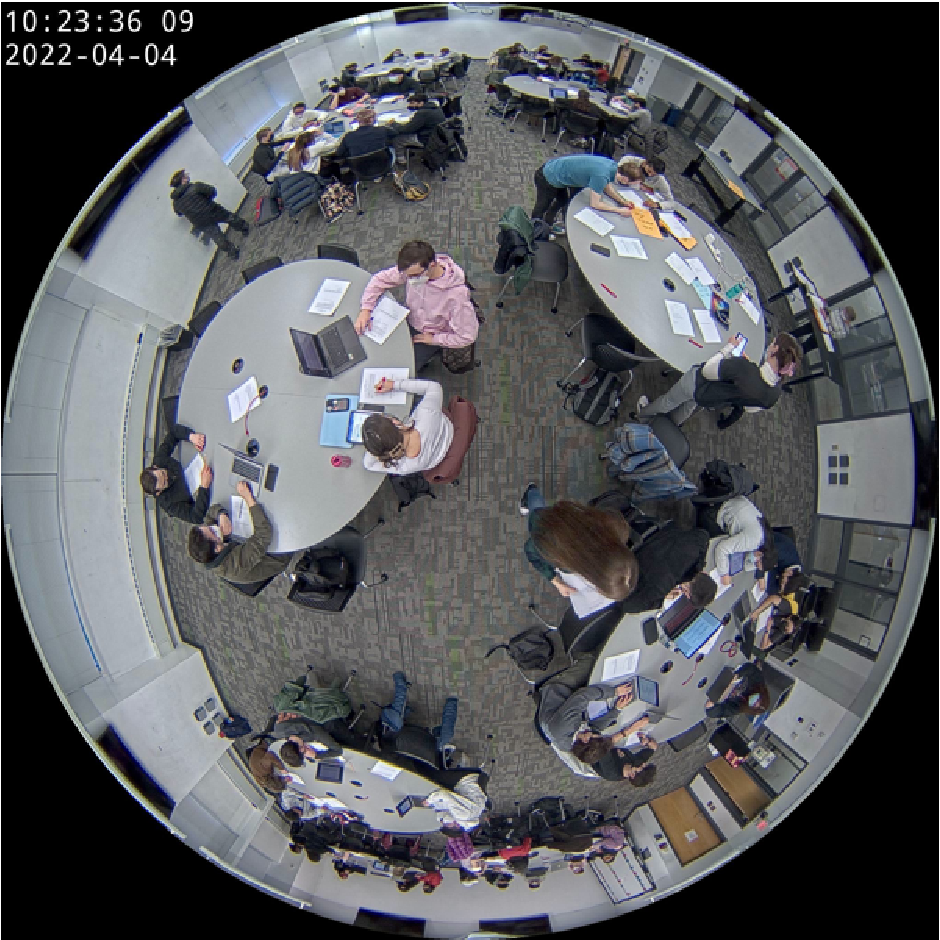,width=6cm}}
  \medskip
  \centerline{(c) High-complexity scenario (images 201-300): 50+ people}
  \caption{Example 2,048$\times$2,048-pixel images used in RAPiD's complexity evaluation.}
  \label{fig:images}
\end{figure}

In order to test RAPiD's complexity as a function of varying occupancy scenarios, we recorded 300 JPEG video frames at 2,048$\times$2,048-pixel resolution in a 2,000 ft$^2$ classroom as follows:
\begin{itemize}
	\item medium-complexity scenario (images 1-100): 14 people spread out throughout the space (Fig.~\ref{fig:images}(a)),
	\item low-complexity scenario (images 101-200): 5-6 people in the space (Fig.~\ref{fig:images}(b)),
	\item high-complexity scenario (images 201-300): 50+ people widely spread out throughout the space (Fig.~\ref{fig:images}(c)).
\end{itemize}

\begin{figure}[htb]
  \centerline{\epsfig{figure=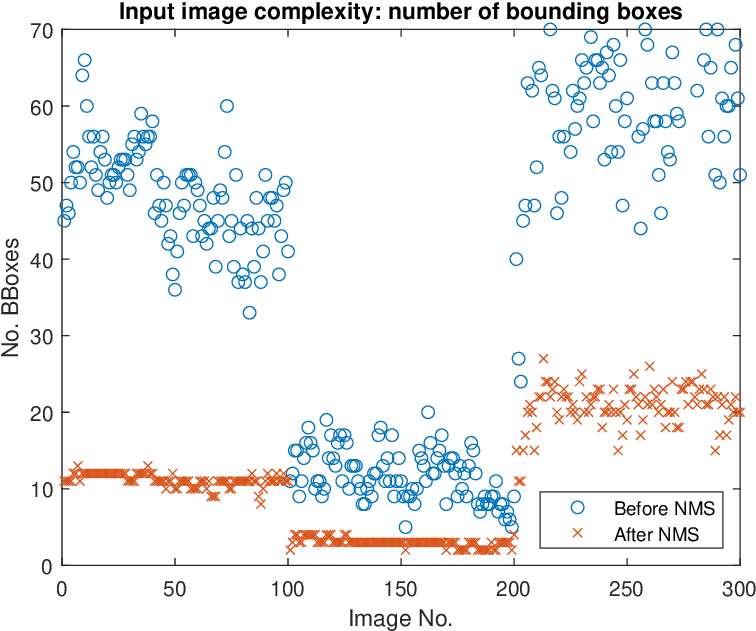,width=8.0cm}}	
  \caption{Scene complexity: medium (images 1-100), low (images 101-200), and high (images 201-300). The blue circles denote the number of bounding boxes detected by RAPiD before applying NMS (non-maximum suppression), while the orange crosses denote the number of bounding boxes remaining after NMS.}
  \label{fig:data_bboxes}
\end{figure}

Fig.~\ref{fig:data_bboxes} shows the number of bounding boxes detected by RAPiD before and after non-maximum suppression (NMS). Clearly, the fewest number of bounding boxes are produced for images 101-200 since only 5-6 people are present.

\section{Experimental results}

The 8559U CPU in the NUC consists of 4 physical cores and 2 threads running on each core, for the total of 8 threads. Threads are referred to as ``logical cores'' in the Windows operating system. The PyTorch library used in RAPiD supports hyperthreading, and we will refer to those threads as ``PyTorch threads''.  We study three mechanisms of logical-core/thread allocation:
\begin{itemize}
	\item Ubuntu-managed,
	\item user-managed via system calls (assignment of logical cores),
	\item user-managed via PyTorch-library calls (limitation of the number of PyTorch threads).
\end{itemize}


\subsection{Ubuntu-managed logical-core allocation}
\label{ssec:ubuntu}

In the first set of experiments, we allowed Ubuntu to manage computing resources, i.e., assign tasks to logical cores (threads).


In the experiments, we measured the following system characteristics:
\begin{itemize}
	\item inference time for a single image,
	\item percentage load for all threads (logical cores),
	\item temperature of the CPU and physical cores,
	\item clock frequency of the CPU.
\end{itemize}  

Fig.~\ref{fig:rapid1} shows system performance for a single instance of RAPiD running on the CPU. The average inference time is 1.8sec and the system load is multiplexed between the 8 logical cores (threads). The CPU and individual core temperatures are correlated with scene complexity and vary much less for images 101-200 that include only 5-6 people. Also, there is much less variation of thread load for images 101-200 (low-complexity scenario) than for other images; the load drops to zero much less frequently (fewer large discontinuities in the load plot). This is likely due to fewer bounding boxes detected in images 101-200. Also, the CPU clock frequency drops for images 101-200. However, the inference time is fairly constant (in the range of 1.66-1.89sec with standard deviation of 0.03) regardless of complexity. It is clear from the inference-time plot, that the system spends almost all of the time on detecting bounding boxes (red line). Image pre-processing (resizing from 2,048$\times$2,048 to 1,024$\times$1,024 and conversion to tensor dimensions shown as a blue line) and non-maximum suppression (NMS shown as a yellow line) take less than 0.1sec together, thus not affecting the overall inference time in any significant way.

Fig.~\ref{fig:rapid2} shows system performance for two instances of RAPiD running simultaneously. At the top of the left column we replaced the plot of times taken by individual components of RAPiD (pre-processing, bounding-box detection, NMS and total times shown in Fig.~\ref{fig:rapid1}) with two plots of the total time taken by each of the two instances of RAPiD running in parallel. The average inference time for 2 images is 3.23sec (computed over all images) and both RAPiD instances show almost identical inference times for the same images. The standard deviation is also low at about 0.04sec with all inference times in the range of 3.09-3.33sec. At the top of the right column we replaced the plot of the number of bounding boxes with a plot of the maximum time it takes to process a pair of images in parallel (the larger time of the two times needed to complete the two images by two instances of RAPiD running simultaneously). The average, standard deviation and min/max values of these maximum times computed over all 300 image pairs are very similar to the corresponding values from the top-left plot, suggesting that all image pairs are processed almost synchronously. The computing load is evenly distributed between the 8 logical cores (threads) and there is no task switching from one thread to another (that is ending a task in one thread and starting in another, thus creating large discontinuities in the load plot). Such switching occurred for a single RAPiD instance shown in Fig.~\ref{fig:rapid1}. We believe this may explain the fact that the 3.23sec needed to process 2 images in parallel is less than the time to process them sequentially by a single instance of RAPiD (2$\times$1.8sec = 3.6sec). The temperature varies less for the 2-instance RAPiD than for the single run and its average settles slightly below 80$^\circ$C. Interestingly, the CPU clock frequency settles at slightly above 3GHz after about half of the images have been processed by the 2-instance RAPiD but remains at about 4GHz for the single RAPiD.

Fig.~\ref{fig:rapid3} shows system performance for three instances of RAPiD running simultaneously. The top two plots correspond to those from Fig.~\ref{fig:rapid2} but for the case of 3 instances of RAPiD. The average inference time is now 6.03sec for 3 images, more than 3$\times$1.8sec = 5.4sec if one were to run RAPiD sequentially. This is due to heavy CPU load as all 8 logical cores (threads) are running at about 100\%. The temperature slightly exceeds 80$^\circ$C and the clock frequency stays slightly above 3GHz. Note that the standard deviation of the inference times is now higher at about 0.2sec and the range is 1.93-7.3sec. The minimum time of 1.93sec happens at the very end of processing when two images of the last triplet have been already completed and all CPU resources are available to process a single image. However, the maximum time of 7.3sec and many instances close to 7sec are due to the competition for resources. This inference time variability is a potential problem since in order to keep images (and people detections) time-synchronized, the system would have to wait until all 3 images have been completed rather than processing them as fast as possible. The plot of the maximum inference time for each image triplet is shown at the top of the right column. On average, the system would have to wait 6.38sec before acquiring another image triplet, with the minimum of 5.68sec and maximum of 7.3sec. 

The performance for 4 instances of RAPiD running in parallel is shown in Fig.~\ref{fig:rapid4}. The average inference time in asynchronous processing (top-left plot) is 9.92sec for 4 images and there is a significant inference time variation for different images. The average of the maximum inference time (synchronous processing) is 11.4sec, quite a bit higher than the average asynchronous-processing time of 9.92sec. The thread load is at 100\% and there is a load drop-off at the end since images are not processed in sync. The CPU temperature slightly increased to about 82-84$^\circ$C and the clock frequency increased to about 3.3GHz.

For 5 instances of RAPiD running in parallel (Fig.~\ref{fig:rapid5}), the average inference time in asynchronous processing is 14.5sec for 5 images and there is a large inference-time variation. The average of the maximum time to synchronously process 5 images is 16.3sec, but it can be as high as 20.5sec. The thread load is 100\%. The CPU temperature is about 82-84$^\circ$C and the clock frequency has increased to about 3.5GHz.

Instead of plotting results for 6-8 instances of RAPiD running in parallel, we are summarizing the main statistics in Table~\ref{tbl:rapid_x_times}. Clearly, in order to synchronously process all images in an $N$-tuple, more time is needed than when processing asynchronously (except for a single instance of RAPiD). The average time increase ranges from 0.01sec for RAPiD x 2 to 1.48sec for RAPiD x 4 to 1.40sec for RAPiD x 8. Shown in a boldface font are effective average times needed to synchronously process $N$ images in parallel.

\renewcommand{\arraystretch}{1.2}
\begin{table*}[htb]
	\centering
	\caption{Statistics of inference times in seconds for a varying number of RAPiD instances running in parallel. The ``$N$-image inference time'' statistics are obtained by first computing the average inference time for each $N$-image tuple and then computing the average and standard deviation of these 300 average times. The ``$N$-image maximum inference time'' statistics are obtained by first computing the {\it maximum} inference time for each $N$-image tuple and then computing the average, standard deviation, minimum and maximum of these 300 maximum times. \label{tbl:rapid_x_times}}
	\medskip
	\begin{tabular}{c|cc|ccccc}
		\hline
		& \multicolumn{2}{c|}{$N$-image inference time} & \multicolumn{4}{c}{$N$-image maximum inference time}\\
		RAPiD x $N$ & Average & Std.\ dev. & {\bf Average} & Std.\ dev. & Min & Max \\
		\hline
		RAPiD x 1 & 1.80 & 0.03 & {\bf 1.80} & 0.00 & 1.66 & 1.89 \\
		RAPiD x 2 & 3.23 & 0.04 & {\bf 3.24} & 0.04 & 3.11 & 3.33 \\
		RAPiD x 3 & 6.03 & 0.21 & {\bf 6.38} & 0.28 & 5.68 & 7.30 \\
		RAPiD x 4 & 9.92 & 0.85 & {\bf 11.4} & 1.15 & 8.42 & 15.4 \\
		RAPiD x 5 & 14.5 & 0.97 & {\bf 16.3} & 1.11 & 12.9 & 20.5 \\
		RAPiD x 6 & 19.2 & 0.72 & {\bf 21.0} & 0.88 & 18.8 & 24.2 \\
		RAPiD x 7 & 21.6 & 0.65 & {\bf 23.1} & 0.67 & 21.6 & 25.0 \\
		RAPiD x 8 & 23.3 & 0.50 & {\bf 24.7} & 0.60 & 23.3 & 26.3 \\
		\hline
	\end{tabular}
\end{table*}

\subsection{User-managed logical-core allocation}
\label{ssec:aff}

In the second set of experiments, the user assigns a logical core, or a set of logical cores, to run an instance of RAPiD inference on. The software-to-hardware allocation was implemented using the Python library function {\tt os.sched\_setaffinity(pid,mask)}, where {\tt pid} is the process ID number and {\tt mask} is a list of logical cores. For example, for {\tt mask = \{0,1\}} RAPiD instance specified by {\tt pid} will only be executed on logical cores \#0 and \#1.


Fig.~\ref{fig:rapid0} shows the results when a single instance of RAPiD is limited to run on a single logical core (\#0). The inference time is 23.8sec with standard deviation of 0.04sec. While logical core (thread) \#0 is 100\% occupied (blue line in the middle-left plot), only logical core \#1 is slightly used (likely by the operating system). Logical cores \#2-7 are completely idle. The CPU frequency is high at about 4.3GHz.

When a single instance of RAPiD is allowed to use two logical cores (\#0 and \#1) for inference, the average inference time drops to 11.7sec (Fig.~\ref{fig:rapid01}) with standard deviation of 0.09sec. Only logical cores \#0 and \#1 are under heavy load. A similar inference time of 11.2sec and standard deviation of 0.08sec are observed for 3 logical cores used (Fig.~\ref{fig:rapid012}). However, the average inference time drops to 1.82sec when 4 logical cores are allowed (\#0, 1, 2, 3), shown in Fig.~\ref{fig:rapid0123}. Increasing the number of logical cores to 5, 6, 7 (Figs.~\ref{fig:rapid01234}-\ref{fig:rapid0123456}) keeps the inference time around 1.8sec and standard deviation around 0.02-0.03sec. Therefore, there is no benefit to using more than 4 logical cores when executing a single instance of RAPiD.

We summarize the average execution times for running a single instance of RAPiD on different hardware allocations in Table~\ref{tbl:aff1_times}. Clearly, at least 4 logical cores need to be allocated to RAPiD for best performance.
\renewcommand{\arraystretch}{1.2}
\begin{table*}[htb]
	\centering
	\caption{Average inference times in seconds for a single instance of RAPiD and different hardware allocations (logical cores 0-7).\label{tbl:aff1_times}}
	\medskip
	\begin{tabular}{c|ccccccc}
		\hline
		& \multicolumn{7}{c}{Logical cores allowed}\\
		& 0 & 0--1 & 0--2 & 0--3 & 0--4 & 0--5 & 0--6\\\hline
		RAPiD x 1 & 23.75 & 11.69 & 11.24 & 1.82 & 1.84 & 1.81 & 1.80\\
		\hline
	\end{tabular}
\end{table*}

After timing a single instance of RAPiD inference with a varying number of logical cores, the experiment was repeated with 2, 4, or 8 instances of RAPiD running in parallel. Each instance of RAPiD was assigned its own, private logical cores, restricted from being shared with other instances. 
The average execution times are shown in Table~\ref{tbl:affn_times}. When running 2 instances of RAPiD, one instance runs on logical cores \#0-3, while the other instance runs on logical cores \#4-7. Similarly, for 4 instances of RAPiD they run respectively on logical cores \#0-1, \#2-3, \#4-5, \#6-7. Finally, in the case of 8 instances of RAPiD each runs on its own logical core.
\begin{table*}[htb]
	\centering
	\caption{Average inference times in seconds for 2, 4 and 8 instances of RAPiD running in parallel for different hardware allocations (logical cores 0-7). See the caption of Table~\ref{tbl:rapid_x_times} for the explanation of statistics calculation. \label{tbl:affn_times}}
	\medskip
	\begin{tabular}{l|cc|cc}
		\hline
		RAPiD x $N$ (assigned cores) & \multicolumn{2}{c|}{$N$-image inf.\ time} & \multicolumn{2}{c}{$N$-image max inf.\ time} \\
		& Average & Std.\ dev.\ & {\bf Average} & Std.\ dev. \\
		\hline
		RAPiD x 2 (0-3/4-7)         & 3.19  & 0.04 & {\bf 3.20} & 0.03 \\
		RAPiD x 4 (0-1/2-3/4-5/6-7) & 14.25 & 0.39 & {\bf 14.50} & 0.37 \\
		RAPiD x 8 (0/1/2/3/4/5/6/7) & 26.68 & 0.99 & {\bf 27.89} & 0.77 \\
		\hline
	\end{tabular}
\end{table*}

As expected, the best performance is obtained by running 2 instances of RAPiD, each on 4 logical cores, which requires 3.19sec to complete 2 images on average. This is consistent with the time of 3.23sec from Table~\ref{tbl:rapid_x_times} when Ubuntu handles all threads. In this case, it seems immaterial whether Ubuntu dynamically allocates logical cores or if each instance of RAPiD is manually restricted to a separate set of 4 logical cores.

The average inference time when running 4 instances of RAPiD, each manually-restricted to two logical cores, is 14.25sec for 4 images. This is higher than the 9.92sec reported in Table~\ref{tbl:rapid_x_times} where Ubuntu manages hardware assignments. Clearly, manually restricting each RAPiD instance to a set of logical cores is less efficient than allowing Ubuntu to manage the logical-core assignment. Looking at Table~\ref{tbl:aff1_times}, it is clear that a single instance of RAPiD restricted to 2 logical cores (\#0-1) results in 11.69sec inference time per image. Theoretically, running another 3 instances of RAPiD on the remaining cores one should be able to complete all 4 images in 11.69sec but in practice the operating system runs its own tasks and takes away resources, thus increasing RAPiD's inference time to over 14sec.

When running 8 instances of RAPiD, each restricted to a single logical core, it takes 26.68sec to process 8 images. This is again higher than the result reported in Table~\ref{tbl:aff1_times}, where a single-core execution takes 23.75sec. Again, system tasks slow down the inference.

In order to synchronously process $N$ = 2, 4 or 8 images, the maximum inference time per each $N$-tuple is important. As can be seen in Table~\ref{tbl:affn_times}, for 2 images processed in parallel there is no additional time-cost penalty for synchronous processing (the average of the maximum inference time of 3.20sec is very close to the overall average of 3.19sec). For 4 images processed in parallel, there is a slight penalty of 0.25sec, and for 8 images - a penalty of over 1sec.


\subsection{User-managed PyTorch-thread allocation}
\label{ssec:thread}

In the third set of experiments, the user controls the number of threads that PyTorch is allowed to use. This is accomplished by means of {\tt torch.set\_num\_threads(N)} function call. This function sets the number of threads {\tt N} that can be used for intra-operation parallelism on the CPU. 
No hardware (logical core) limitations are applied at the system level, as was done in Section~\ref{ssec:aff}. Note that we were unable to obtain results for the case of one PyTorch thread ({\tt torch.set\_num\_threads(1)}). The system kept crashing after a few images without a clear indication of the issue. We suspect that this could be due to some outdated software packages, but we refrained from a full Ubuntu update since currently the multi-camera occupancy sensing system works very well and is extensively used for time-critical experiments and we did not want to risk the system's stability at this critical point.

The results are shown in Table~\ref{tbl:thread_times}. When running 1 or 2 instances of RAPiD, the smallest average inference time is obtained when each instance is restricted to 4 PyTorch threads:
\begin{itemize}
	\item 1.81sec for a single instance of RAPiD,
	\item 3.22sec for two instances of RAPiD.
\end{itemize}
%
It seems that RAPiD's use of PyTorch library is optimized for 4 threads. Indeed, when we checked the number of threads spawned by RAPiD when Ubuntu manages load distribution (Section~\ref{ssec:ubuntu}), we noticed that each instance of RAPiD typically spawns 4 threads but drops to 2 threads when the system is busy. 

However, when running 3 or more instances of RAPiD the smallest average inference time results from a restriction to 2 PyTorch threads:
\begin{itemize}
	\item 4.75sec for three instances of RAPiD,
	\item 6.02sec for four instances of RAPiD,
	\item 12.67sec for eight instances of RAPiD.
\end{itemize}
%
Clearly, the PyTorch library also works very well with two threads only. As we pointed out above, we could not run RAPiD by setting the number of PyTorch threads to 1.

\renewcommand{\arraystretch}{1.2}
\begin{table*}[htb]
	\centering
	\caption{Average inference times in seconds for 1-8 instances of RAPiD running in parallel for a different number of PyTorch threads. Shown in boldface is the smallest inference time for the asynchronous processing (left part of the table) and for synchronous processing (right part of the table). See the caption of Table~\ref{tbl:rapid_x_times} for the explanation of statistics calculation. \label{tbl:thread_times}}
	\medskip
	\begin{tabular}{c|cccccc|ccc}
		\hline
		& \multicolumn{6}{c|}{$N$-image inference time (average)} & \multicolumn{3}{c}{$N$-image max.\ inf.\ time}\\\cline{2-7}
		& \multicolumn{6}{c|}{Number of PyTorch threads allowed} & \multicolumn{3}{c}{for the boldfaced results}\\\cline{8-10}
		& 2 & 3 & 4 & 5 & 6 & 7 & Avg. & Std. dev. & Max \\\hline
		RAPiD x 1 & 2.63 & 2.07 & {\bf 1.81} & 2.33 & 2.09 & 1.92 & {\bf 1.81} & 0.03 & 1.88\\
		RAPiD x 2 & 3.24 & 3.67 & {\bf 3.22} & 4.30 & 5.62 & 10.17 & {\bf 3.23} & 0.04 & 3.38\\
		RAPiD x 3 & {\bf 4.75} & 4.99 & 6.00 & 6.53 & 14.46 & 19.34 & {\bf 5.38} & 0.38 & 5.84\\
		RAPiD x 4 & {\bf 6.02} & 6.94 & 9.90 & 16.23 & 21.29 & 23.75 & {\bf 6.11} & 0.05 & 6.24\\
		RAPiD x 5 & {\bf 7.63} & 9.35 & 14.67 & 21.11 & 23.78 & 27.45 & {\bf 8.06} & 0.22 & 8.68\\
		RAPiD x 6 & {\bf 9.20} & 11.81 & 19.19 & 23.42 & 27.90 & 32.66 & {\bf 9.74} & 0.28 & 11.19\\
		RAPiD x 7 & {\bf 11.03} & 14.57 & 21.72 & 26.64 & 32.42 & 36.55 & {\bf 11.77} & 0.39 & 13.25\\
		RAPiD x 8 & {\bf 12.67} & 17.75 & 23.47 & 30.06 & 36.55 & 32.32 & {\bf 13.78} & 0.94 & 19.16\\
		\hline
	\end{tabular}
\end{table*}

We note that when the number of RAPiD instances running in parallel increases, the smallest inference time does not grow linearly. A simple scaling of the 4-thread inference time of 1.81sec for a single instance of RAPiD would have resulted in 3.62sec for two instances of RAPiD, 5.43sec for three instances, 7.24sec for four instances and 14.48sec for eight instances. The actual smallest inference times in each case are, respectively: 3.22sec, 4.75sec, 6.02sec and 12.67sec. This is a significant reduction of inference time likely due to the reduced thread management cost. In fact, if one were to consider the inference time per image processed, the respective times would have been: 3.22/2 = 1.61sec, 4.75/3 = 1.58sec, 6.02/4 = 1.51sec, 12.67/8 = 1.58sec. {This is a marked improvement over the 1.81sec time for single RAPiD.

While the inference times in Table~\ref{tbl:thread_times} are quite consistent and increase with the number of RAPiD instances running in parallel and with the number of PyTorch threads, there is one inconsistency. In the case of 8 RAPiD instances and 7 PyTorch threads the time of 32.32sec is smaller than the times for 8 instances of RAPiD and 6 PyTorch threads (36.55sec) as well as for 7 instances of RAPiD and 7 PyTorch threads (36.55sec). We have run this simulation several times and this inconsistency persists. Somehow, this combination of 8 RAPiD instances and 7 PyTorch threads better leverages the hardware than the other two scenarios. However, this is not a competing option against running several RAPiD instances on 2 PyTorch threads, so we did not devote any additional time to this issue.

As discussed earlier, the ``faster'' instances of RAPiD need to wait until the ``slower'' instances complete their processing in order to assure time-synchronicity of detections. Therefore, one needs to look at the statistics of the sequence of maximum inference times for each $N$-tuple of images being processed, shown on the right of Table~\ref{tbl:thread_times}. For 2 instances of RAPiD running in parallel, the synchronous processing requires, on average, an additional delay of 0.01sec, for 3 instances - 0.63sec, for 4 instances - 0.09 sec and for 8 instances - 1.11sec. Overall, 8 images can be synchronously processed in less than 14sec on average, with a maximum delay of 19.16sec, while 4 images require slightly more than 6sec with a maximum delay of 6.24sec. In applications that can tolerate this level of delay, this NUC PC can support even 8 cameras thus significantly reducing system costs.

\begin{table*}[!h]
	\centering
	\caption{Average inference times in seconds for asynchronous processing when running 2, 4 and 8 instances of RAPiD in parallel for different hardware-management scenarios.\label{tbl:concl1}}
	\medskip
	\begin{tabular}{c|c|c|c}
		\hline
		RAPiD x $N$ & Ubuntu-managed & User-managed & User-managed\\
		& & (logical cores) & (PyTorch threads)\\\hline
		RAPiD x 2 & 3.23 & {\bf 3.19}  & 3.22 \\
		RAPiD x 4 & 9.92 & 14.25 & {\bf 6.02} \\
		RAPiD x 8 & 23.3 & 26.68 & {\bf 12.67} \\
		\hline
	\end{tabular}
\end{table*}

\section{Conclusions}

The effective inference delay when running multiple instances of RAPiD depends on the application scenario. When people detections produced by RAPiD do not need to be time-synchronized (e.g., cameras are installed in different rooms and their images can be processed independently), the delay is smaller. Table~\ref{tbl:concl1} shows inference times for the three hardware-management scenarios considered in this project. When only two instances of RAPiD run in parallel, all three methods produce similar inference times. However, when 4 or 8 instances of RAPiD are executed, limiting the number of PyTorch threads to 2 produces the smallest delay by a large margin (less than half of the delay of the user-managed logical-core assignment). On this particular Intel CPU, two images can be processed in about 3sec, 4 images - in about 6sec and 8 images in less than 13sec.

When time synchronization is required, such as when multiple cameras monitor the same large space and need to collaborate, the inference delays are slightly larger and are shown in Table~\ref{tbl:concl2}. Again, the user-managed PyTorch-thread allocation produces the smallest delays by a large margin.
\begin{table*}[htb]
	\centering
	\caption{Average inference times in seconds for time-synchronized processing when running 2, 4 and 8 instances of RAPiD in parallel for different hardware-management scenarios.\label{tbl:concl2}}
	\medskip
	\begin{tabular}{c|c|c|c}
		\hline
		RAPiD x $N$ & Ubuntu-managed & User-managed & User-managed\\
		& & (logical cores) & (PyTorch threads)\\\hline
		RAPiD x 2 & 3.24 & {\bf 3.20}  & 3.23 \\
		RAPiD x 4 & 11.4 & 14.50 & {\bf 6.11} \\
		RAPiD x 8 & 24.7 & 27.89 & {\bf 13.78} \\
		\hline
	\end{tabular}
\end{table*}

Clearly, in order to assure the smallest inference delay, all three approaches work equally well when 1 or 2 instances of RAPiD need to run in parallel (support for 1 or 2 cameras, respectively). Otherwise, it is recommended to limit the number of PyTorch threads to 2 for best performance. Both of these conclusions apply to the asynchronous and time-synchronized processing.

One should note that 8559U is an 8-th generation Intel CPU. Today's 12-th generation CPUs would definitely produce shorter times. A further reduction of RAPiD's inference time would be possible by employing GPUs but this, unfortunately, would likely increase system cost.

\parskip=0pt
\parsep=0pt
\bibliographystyle{ieee_fullname}

\bibliography{strings,biblio}

\clearpage

\def\plotsize{6.5cm}

\begin{figure*}[!htb]
  \begin{minipage}[t]{0.48\linewidth}
    \centerline{\epsfig{figure=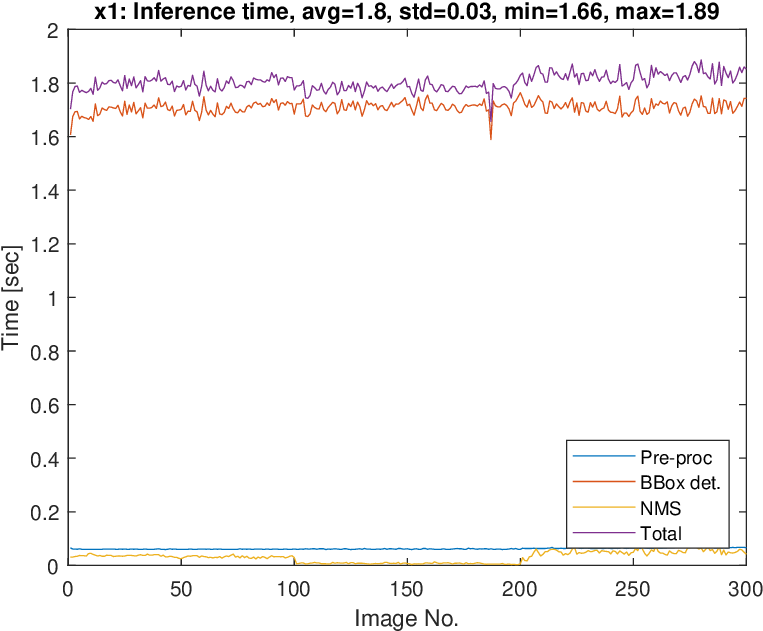,height=\plotsize}}
  \end{minipage}
  \hfill
  \begin{minipage}[t]{0.48\linewidth}
  	\centerline{\epsfig{figure=plots/data_bboxes.eps,height=\plotsize}}
  \end{minipage}
  \bigskip\bigskip
  
  \begin{minipage}[t]{0.48\linewidth}
  	\centerline{\epsfig{figure=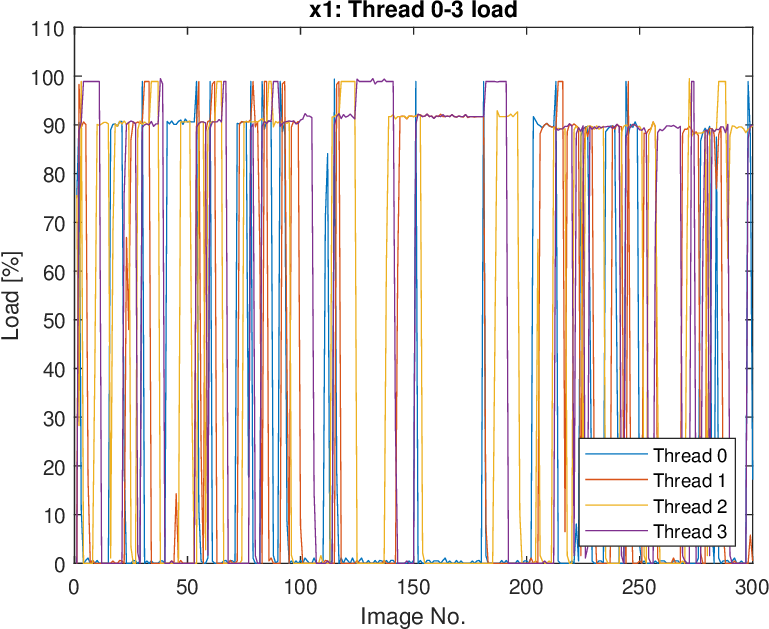,height=\plotsize}}
  \end{minipage}
  \hfill
  \begin{minipage}[t]{0.48\linewidth}
  	\centerline{\epsfig{figure=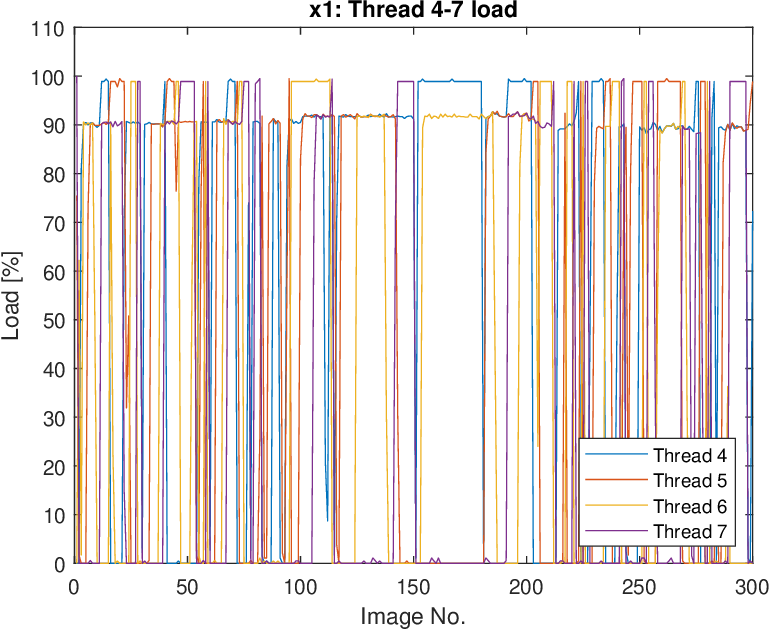,height=\plotsize}}
  \end{minipage}
  \bigskip\bigskip
  
  \begin{minipage}[t]{0.48\linewidth}
  	\centerline{\epsfig{figure=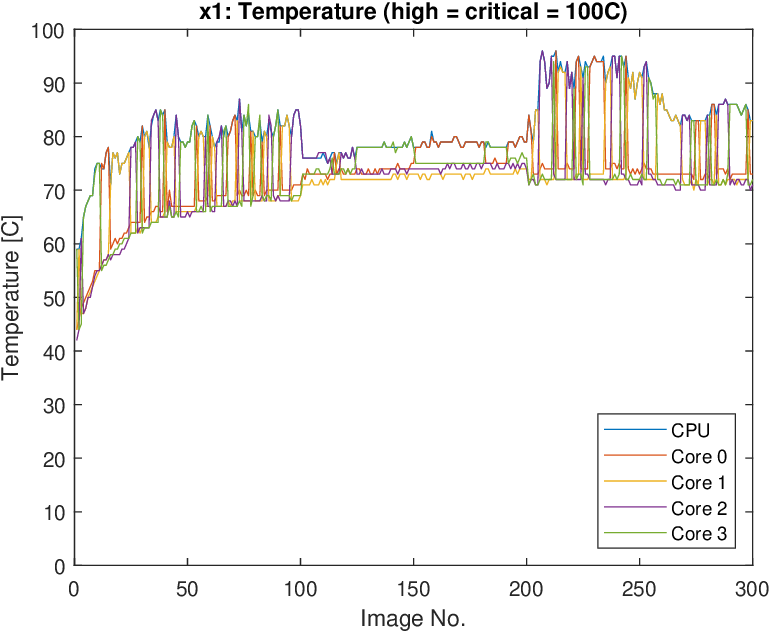,height=\plotsize}}
  \end{minipage}
  \hfill
  \begin{minipage}[t]{0.48\linewidth}
  	\centerline{\epsfig{figure=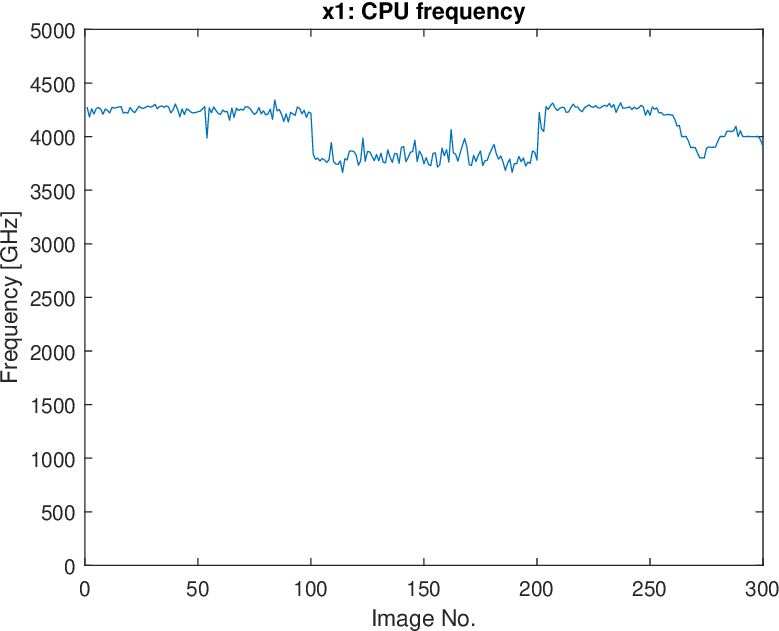,height=\plotsize}}
  \end{minipage}
  \caption{CPU characteristics for 1 instance of RAPiD.}
  \label{fig:rapid1}
\end{figure*}

\begin{figure*}[!htb]
  \begin{minipage}[t]{0.48\linewidth}
	\centerline{\epsfig{figure=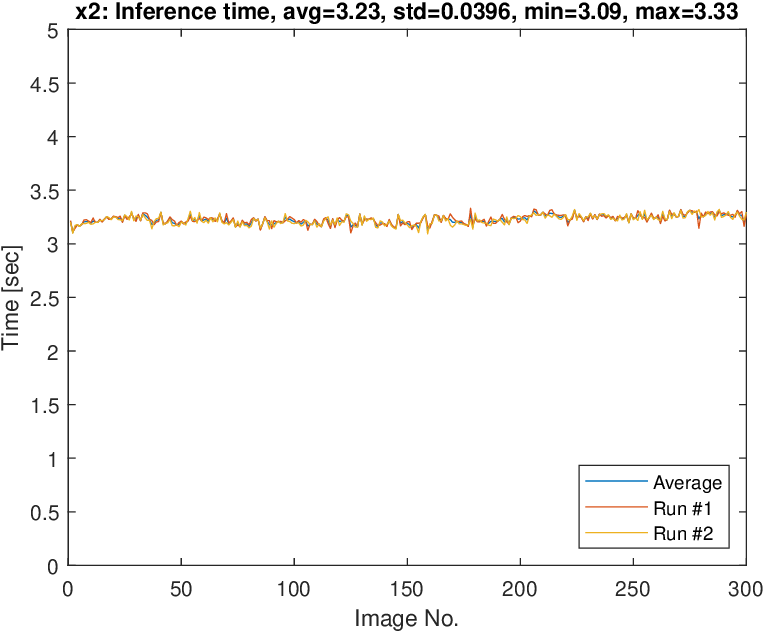,height=\plotsize}}
  \end{minipage}
  \hfill
  \begin{minipage}[t]{0.48\linewidth}
	\centerline{\epsfig{figure=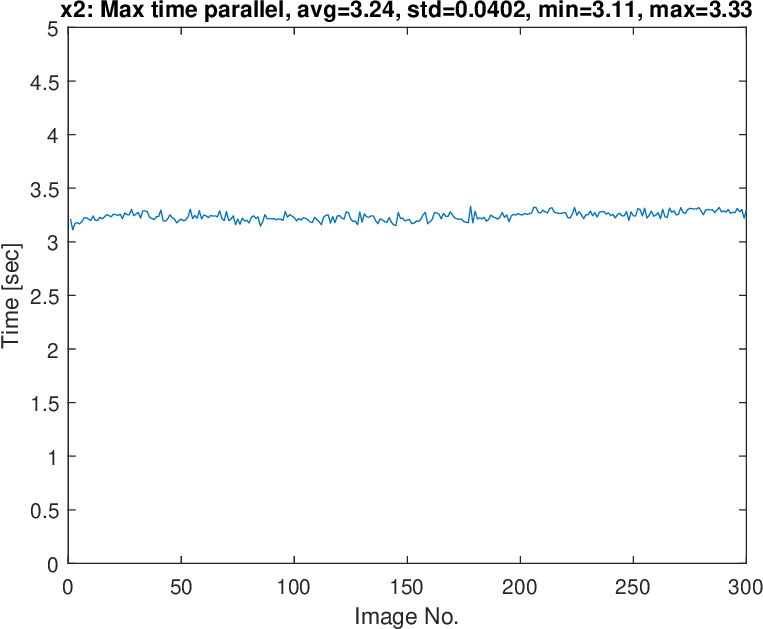,height=\plotsize}}
  \end{minipage}
  \bigskip\bigskip
	
  \begin{minipage}[t]{0.48\linewidth}
	\centerline{\epsfig{figure=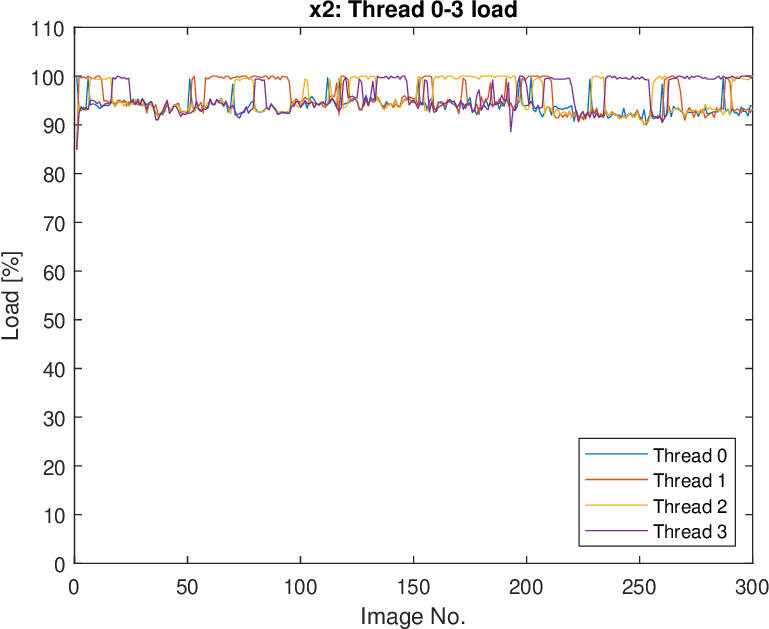,height=\plotsize}}
  \end{minipage}
  \hfill
  \begin{minipage}[t]{0.48\linewidth}
	\centerline{\epsfig{figure=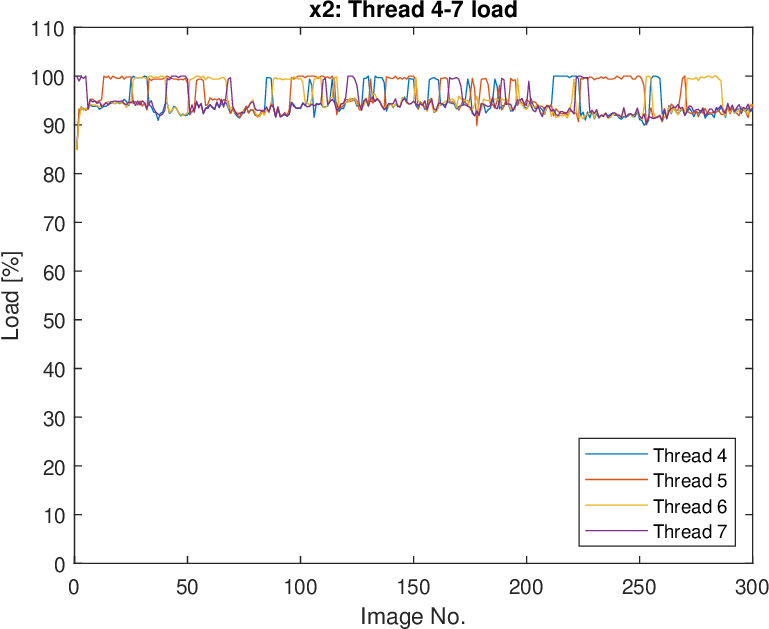,height=\plotsize}}
  \end{minipage}
  \bigskip\bigskip
	
  \begin{minipage}[t]{0.48\linewidth}
    \centerline{\epsfig{figure=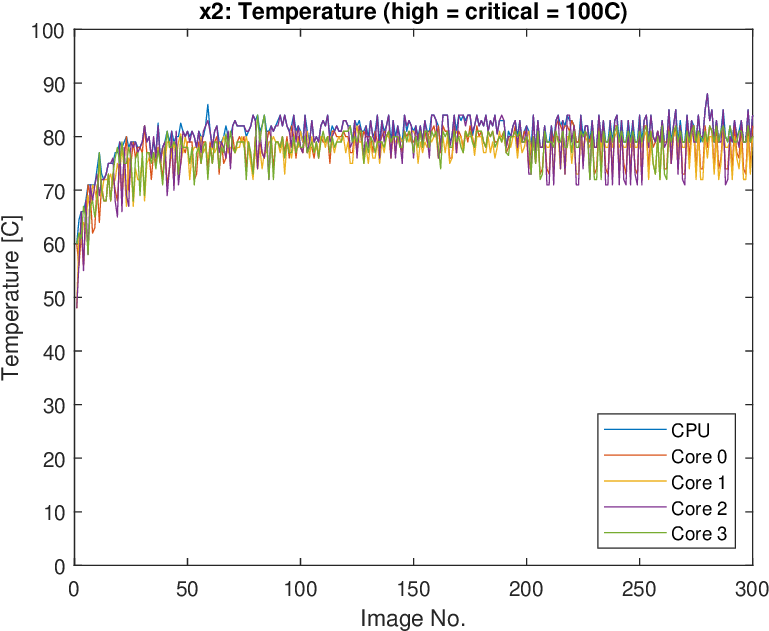,height=\plotsize}}
  \end{minipage}
  \hfill
  \begin{minipage}[t]{0.48\linewidth}
	\centerline{\epsfig{figure=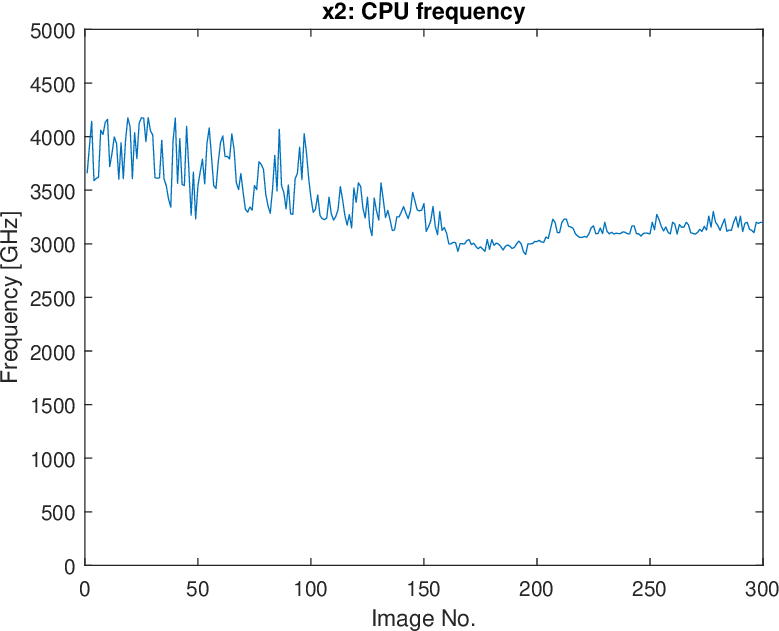,height=\plotsize}}
  \end{minipage}
  \caption{CPU characteristics for 2 instances of RAPiD running in parallel.}
  \label{fig:rapid2}
\end{figure*}

\begin{figure*}[!htb]
  \begin{minipage}[t]{0.48\linewidth}
	\centerline{\epsfig{figure=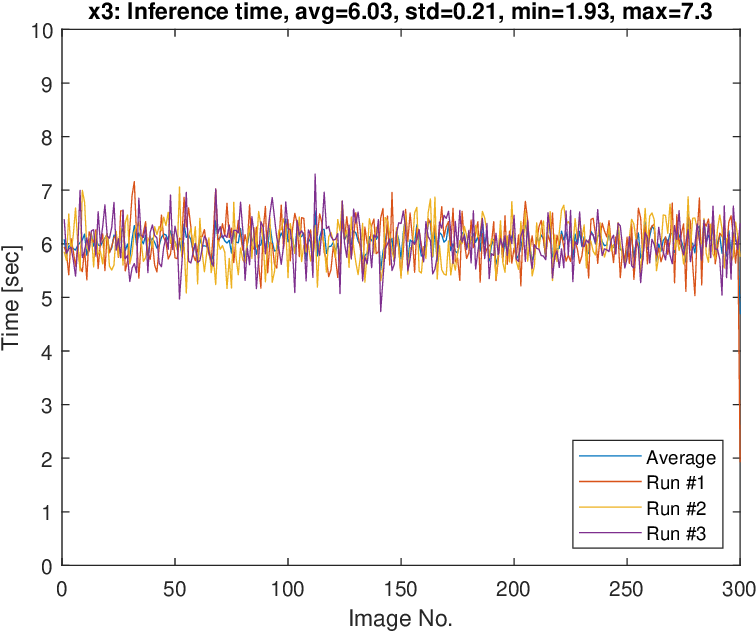,height=\plotsize}}
  \end{minipage}
  \hfill
  \begin{minipage}[t]{0.48\linewidth}
	\centerline{\epsfig{figure=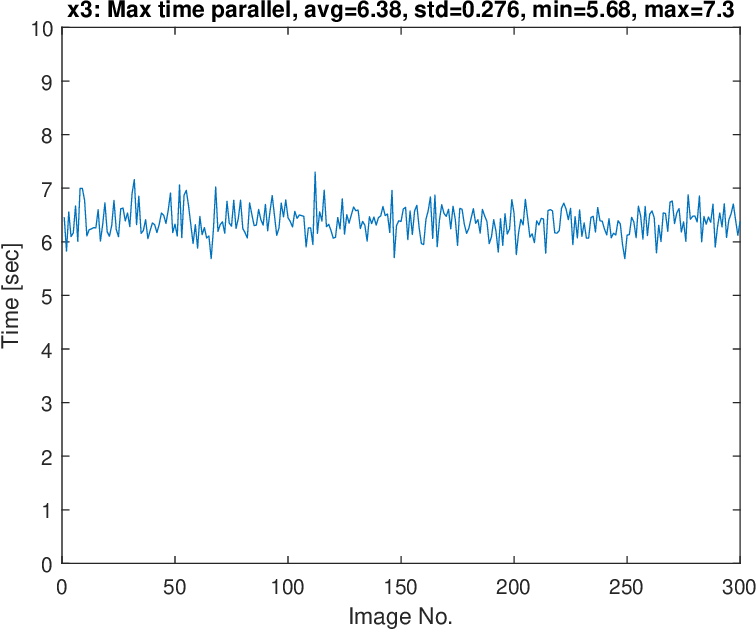,height=\plotsize}}
  \end{minipage}
  \bigskip\bigskip
	
  \begin{minipage}[t]{0.48\linewidth}
	\centerline{\epsfig{figure=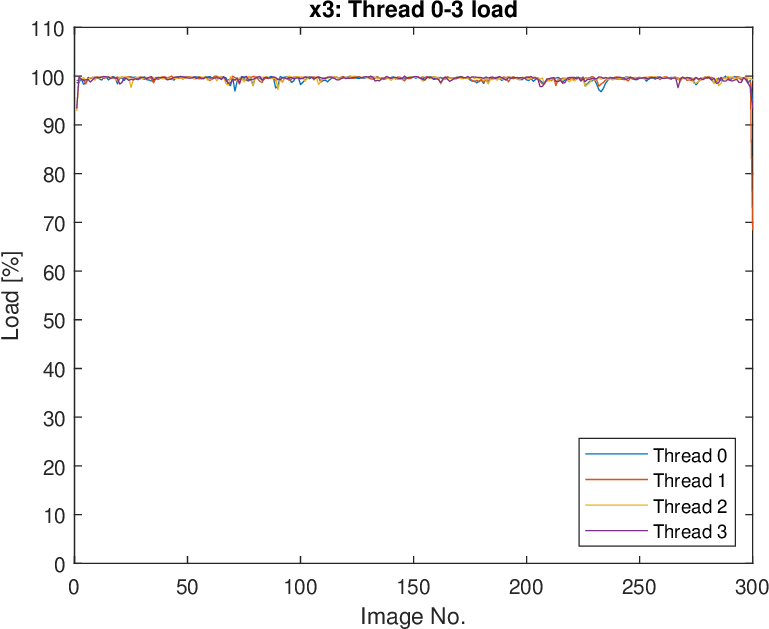,height=\plotsize}}
  \end{minipage}
  \hfill
  \begin{minipage}[t]{0.48\linewidth}
	\centerline{\epsfig{figure=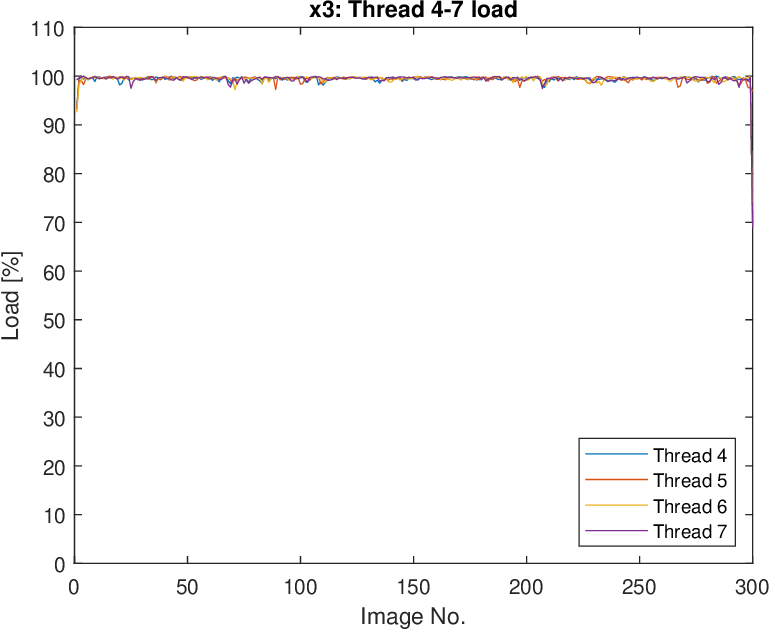,height=\plotsize}}
  \end{minipage}
  \bigskip\bigskip
	
  \begin{minipage}[t]{0.48\linewidth}
	\centerline{\epsfig{figure=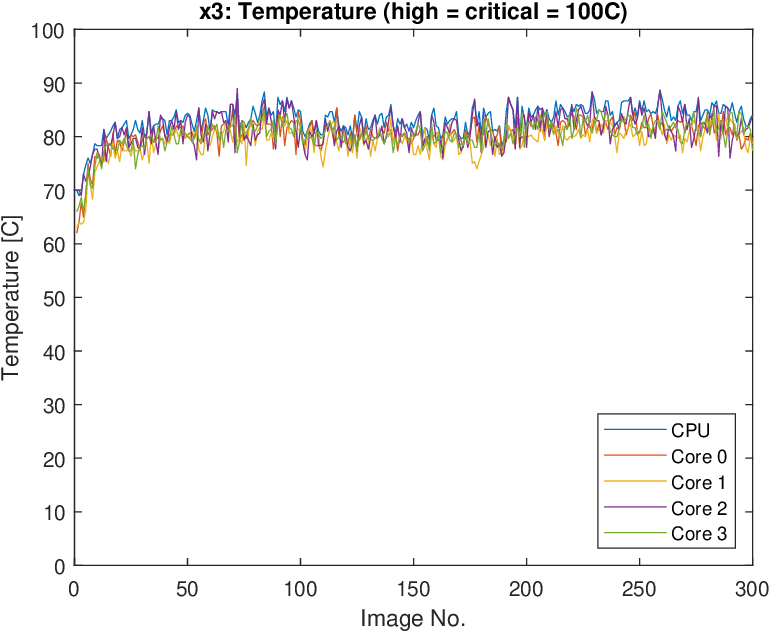,height=\plotsize}}
  \end{minipage}
  \hfill
  \begin{minipage}[t]{0.48\linewidth}
	\centerline{\epsfig{figure=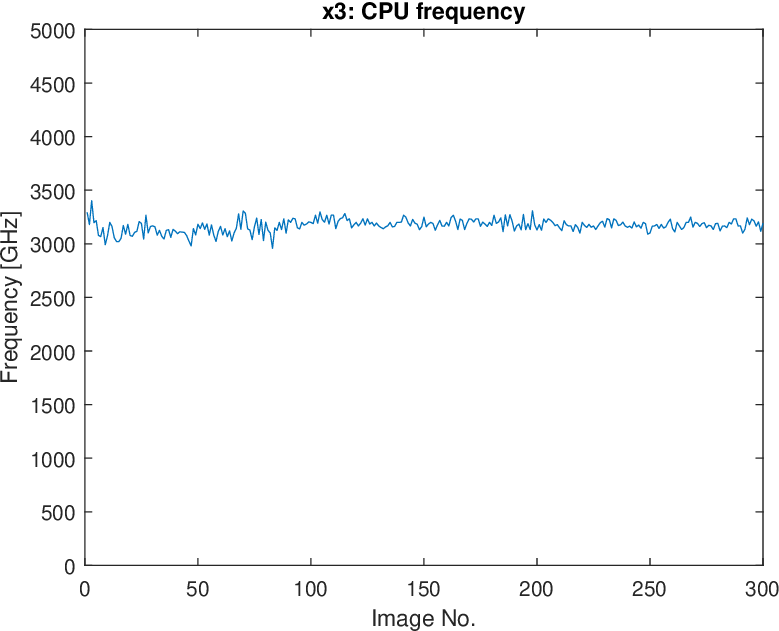,height=\plotsize}}
  \end{minipage}
  \caption{CPU characteristics for 3 instances of RAPiD running in parallel.}
  \label{fig:rapid3}
\end{figure*}

\begin{figure*}[!htb]
  \begin{minipage}[t]{0.48\linewidth}
	\centerline{\epsfig{figure=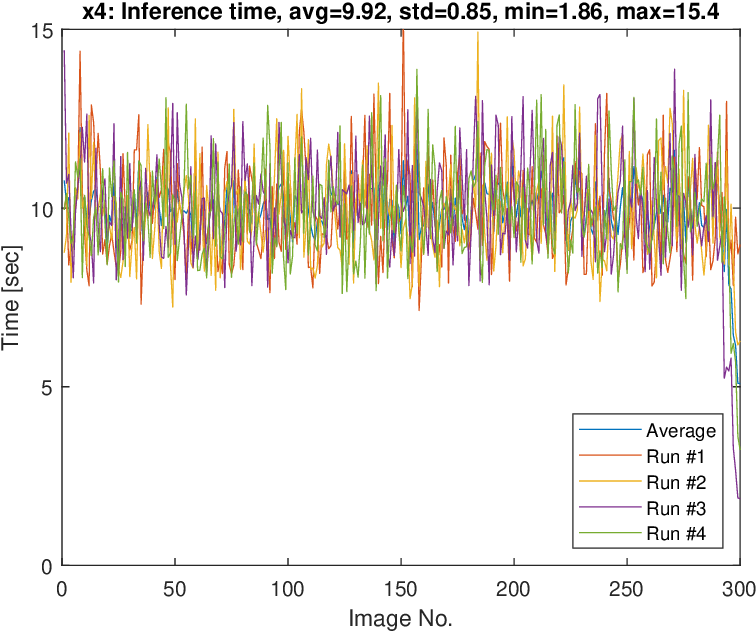,height=\plotsize}}
  \end{minipage}
  \hfill
  \begin{minipage}[t]{0.48\linewidth}
	\centerline{\epsfig{figure=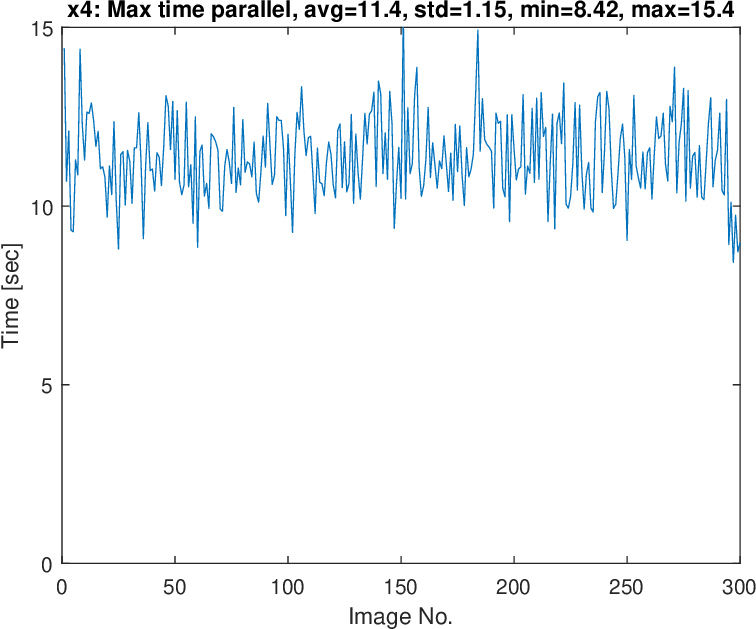,height=\plotsize}}
  \end{minipage}
  \bigskip\bigskip
	
  \begin{minipage}[t]{0.48\linewidth}
	\centerline{\epsfig{figure=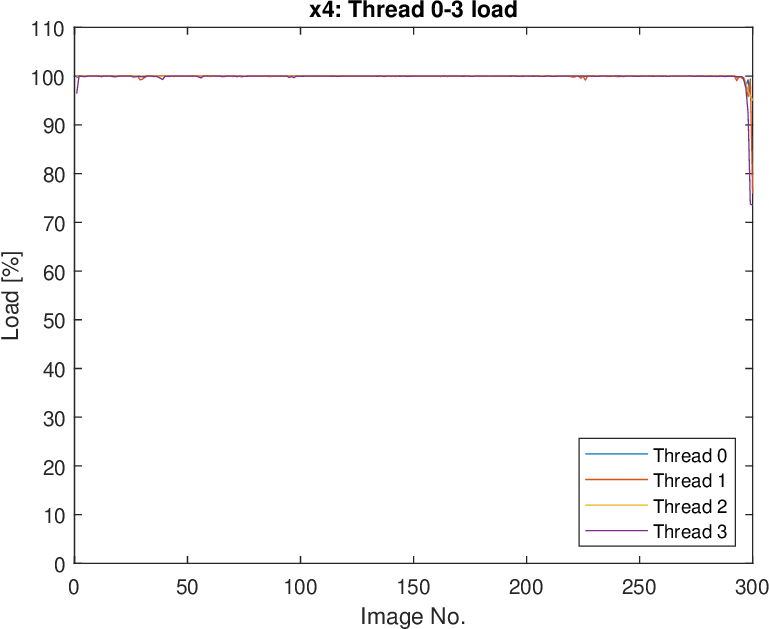,height=\plotsize}}
  \end{minipage}
  \hfill
  \begin{minipage}[t]{0.48\linewidth}
	\centerline{\epsfig{figure=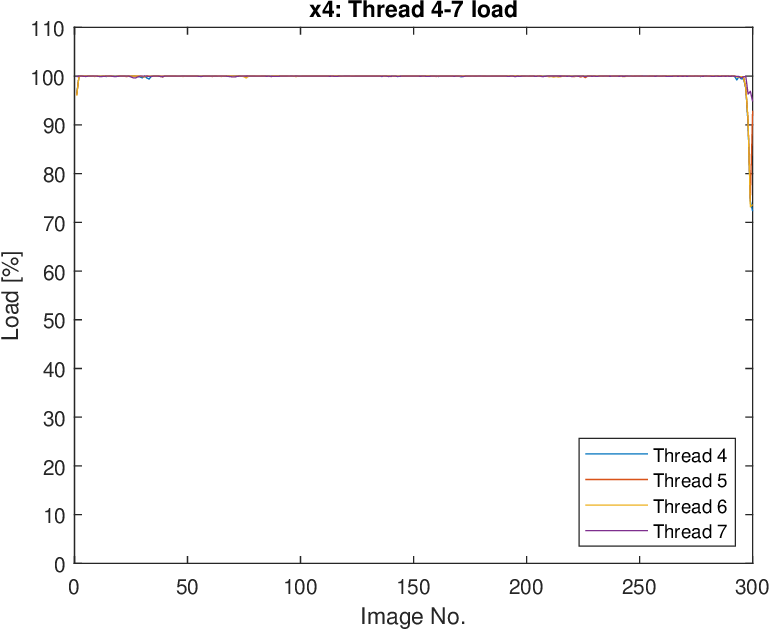,height=\plotsize}}
  \end{minipage}
  \bigskip\bigskip
	
  \begin{minipage}[t]{0.48\linewidth}
	\centerline{\epsfig{figure=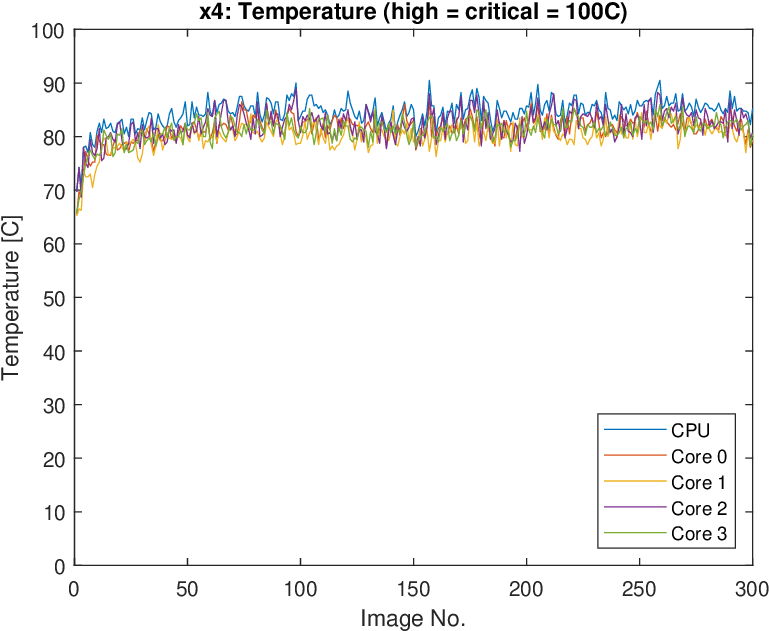,height=\plotsize}}
  \end{minipage}
  \hfill
  \begin{minipage}[t]{0.48\linewidth}
	\centerline{\epsfig{figure=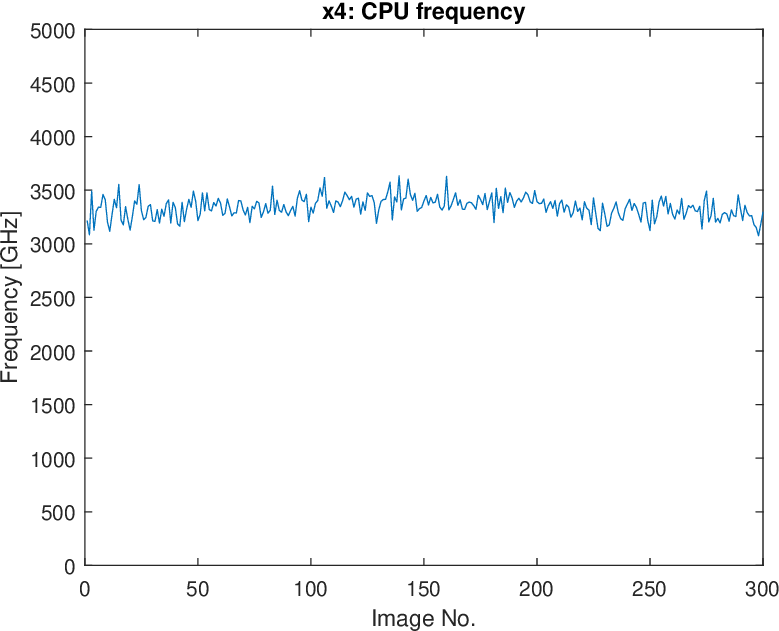,height=\plotsize}}
  \end{minipage}
  \caption{CPU characteristics for 4 instances of RAPiD running in parallel.}
  \label{fig:rapid4}
\end{figure*}

\begin{figure*}[!htb]
  \begin{minipage}[t]{0.48\linewidth}
	\centerline{\epsfig{figure=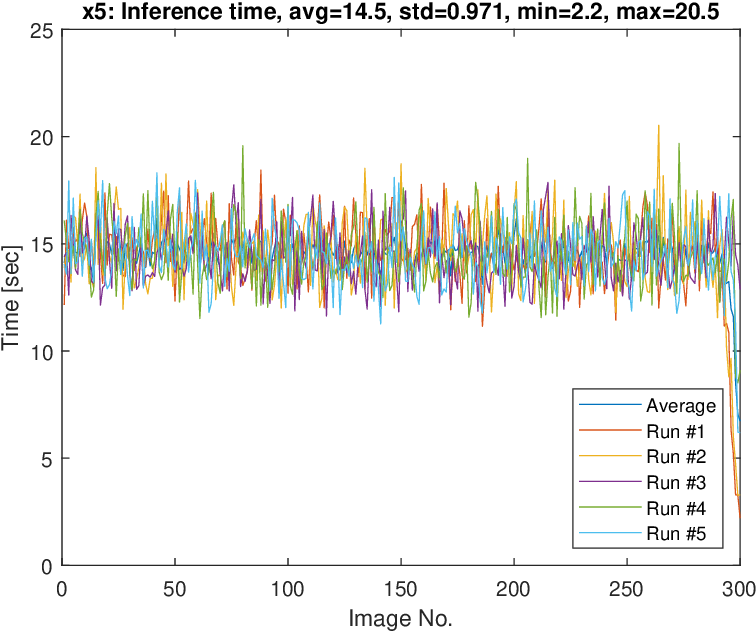,height=\plotsize}}
  \end{minipage}
  \hfill
  \begin{minipage}[t]{0.48\linewidth}
	\centerline{\epsfig{figure=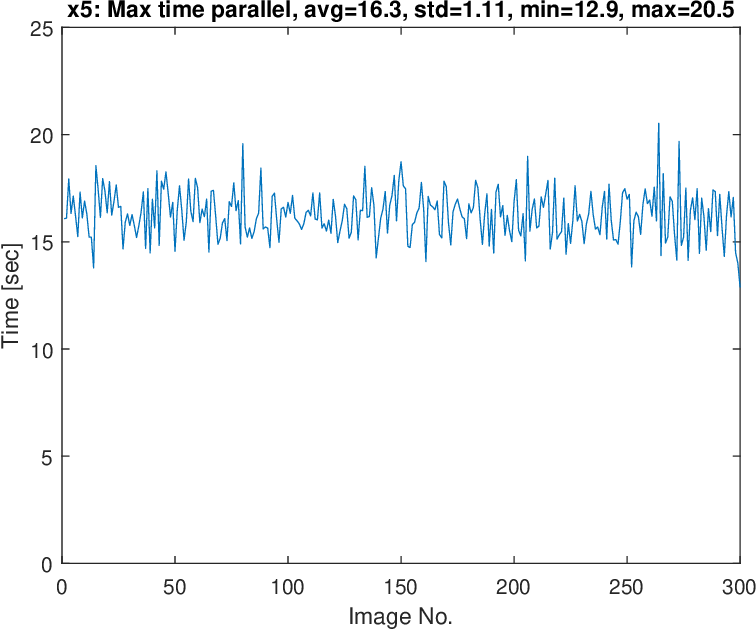,height=\plotsize}}
  \end{minipage}
  \bigskip\bigskip
		
  \begin{minipage}[t]{0.48\linewidth}
	\centerline{\epsfig{figure=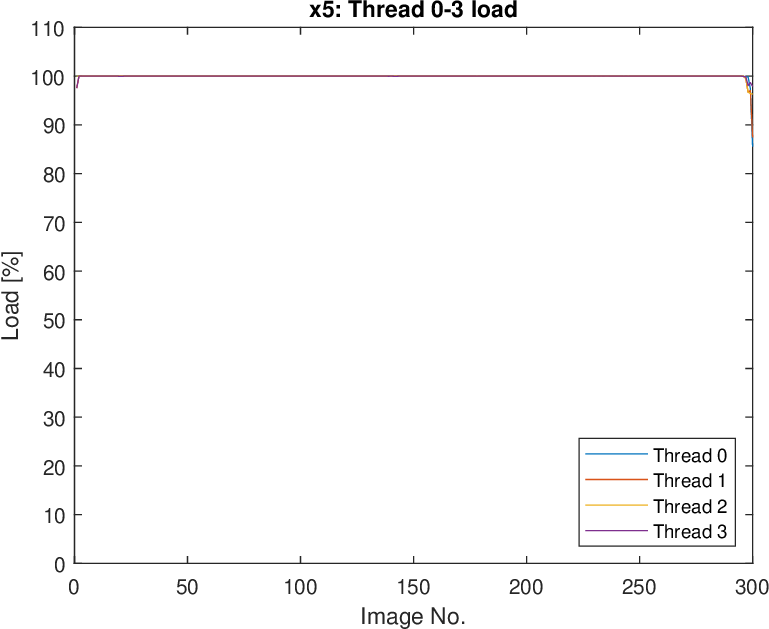,height=\plotsize}}
  \end{minipage}
  \hfill
  \begin{minipage}[t]{0.48\linewidth}
	\centerline{\epsfig{figure=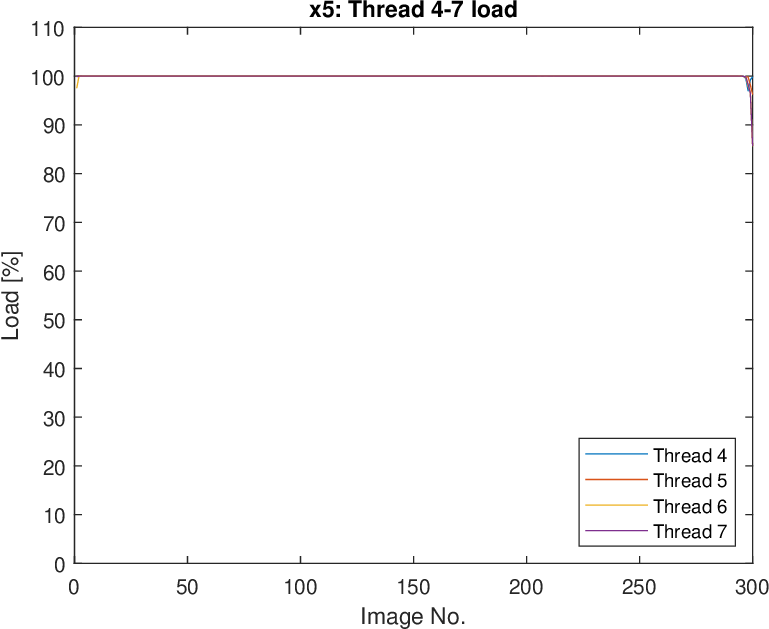,height=\plotsize}}
  \end{minipage}
  \bigskip\bigskip
		
  \begin{minipage}[t]{0.48\linewidth}
	\centerline{\epsfig{figure=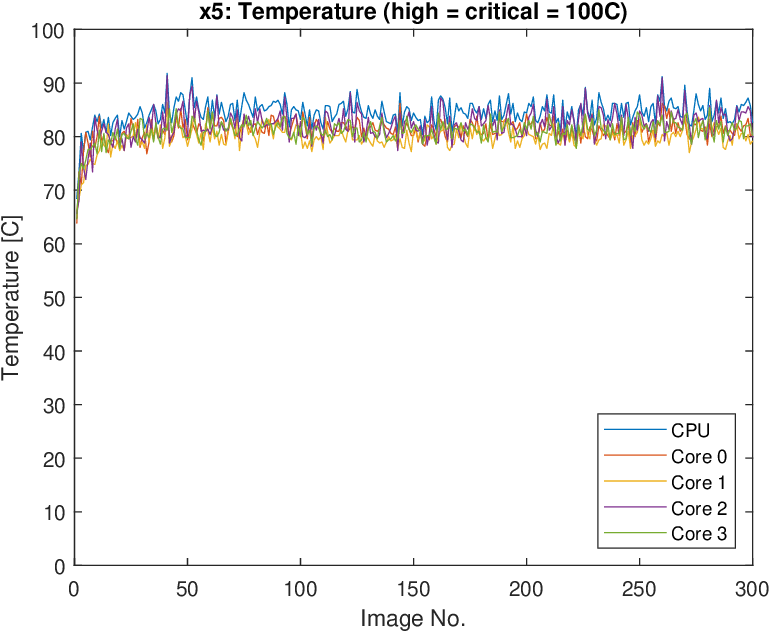,height=\plotsize}}
  \end{minipage}
  \hfill
  \begin{minipage}[t]{0.48\linewidth}
	\centerline{\epsfig{figure=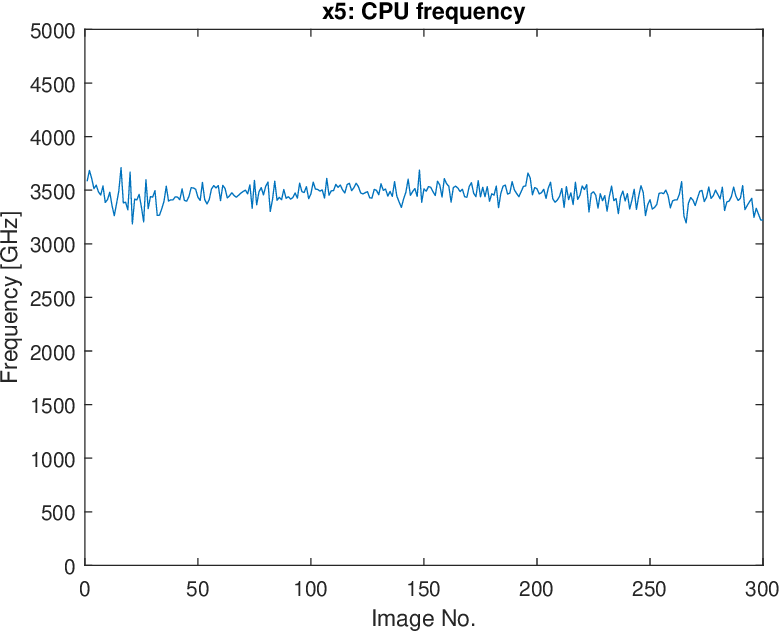,height=\plotsize}}
  \end{minipage}
  \caption{CPU characteristics for 5 instances of RAPiD running in parallel.}
  \label{fig:rapid5}
\end{figure*}

%
%

\begin{figure*}[!htb]
	\begin{minipage}[t]{0.48\linewidth}
		\centerline{\epsfig{figure=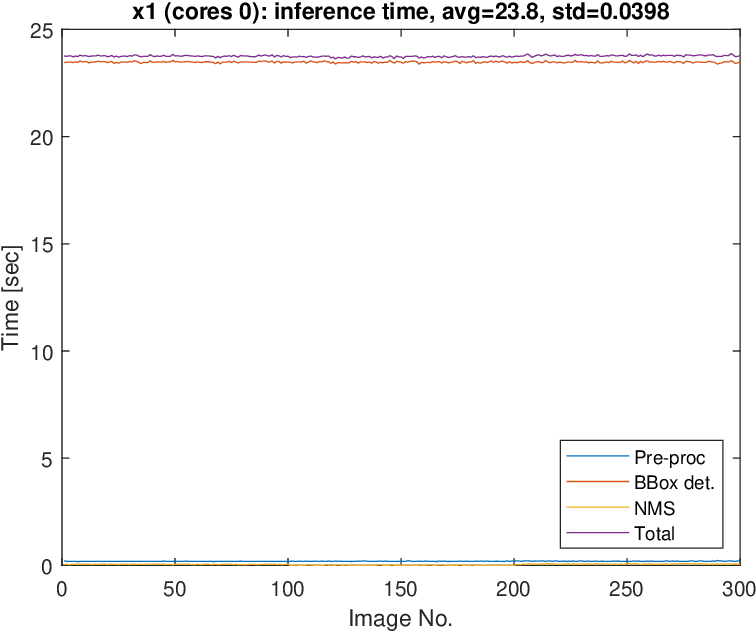,height=\plotsize}}
	\end{minipage}
	\hfill
	\begin{minipage}[t]{0.48\linewidth}
		\centerline{\epsfig{figure=plots/data_bboxes.eps,height=\plotsize}}
	\end{minipage}
	\bigskip\bigskip
	
	\begin{minipage}[t]{0.48\linewidth}
		\centerline{\epsfig{figure=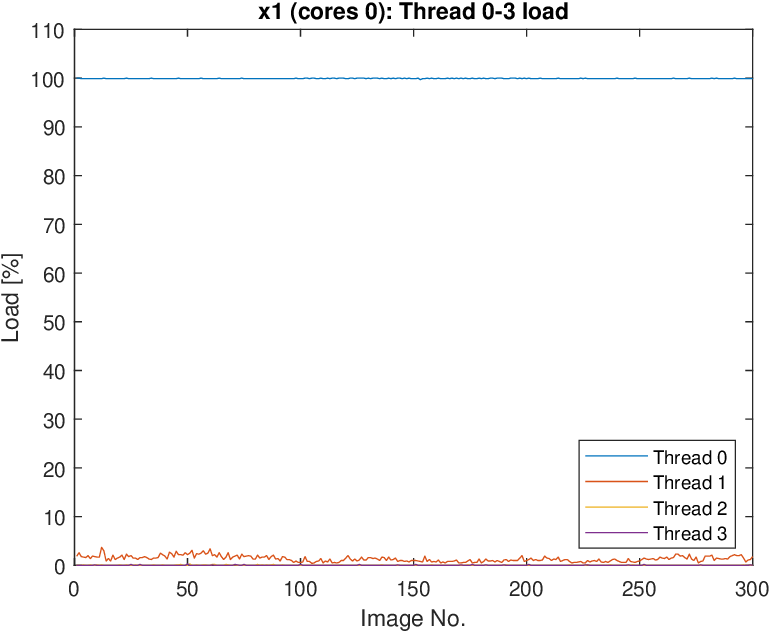,height=\plotsize}}
	\end{minipage}
	\hfill
	\begin{minipage}[t]{0.48\linewidth}
		\centerline{\epsfig{figure=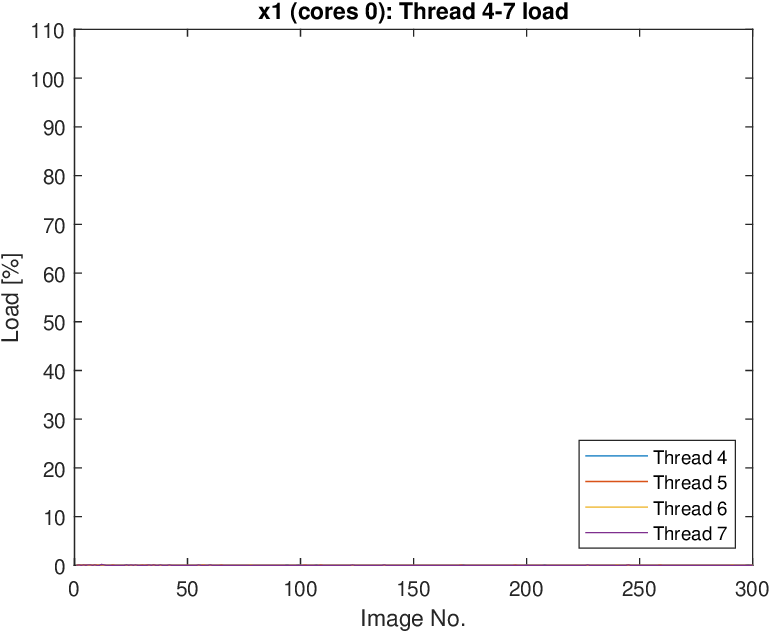,height=\plotsize}}
	\end{minipage}
	\bigskip\bigskip
	
	\begin{minipage}[t]{0.48\linewidth}
		\centerline{\epsfig{figure=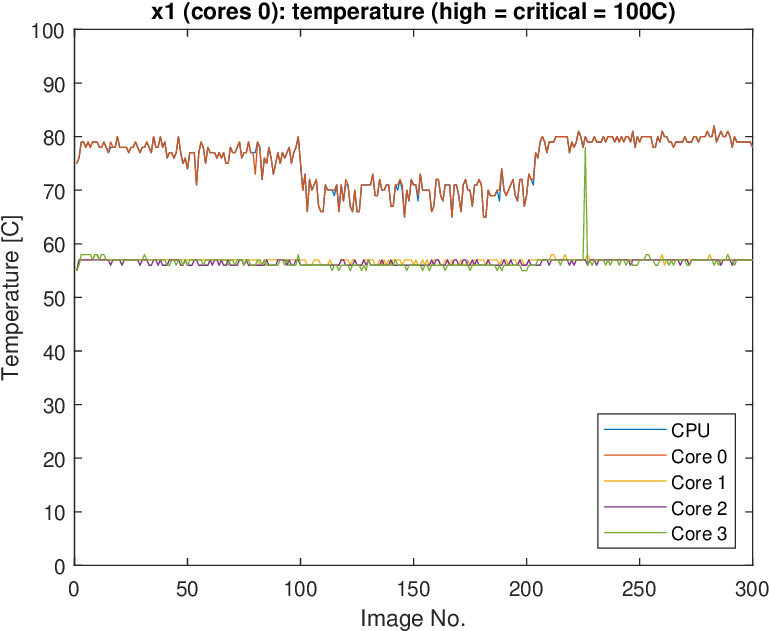,height=\plotsize}}
	\end{minipage}
	\hfill
	\begin{minipage}[t]{0.48\linewidth}
		\centerline{\epsfig{figure=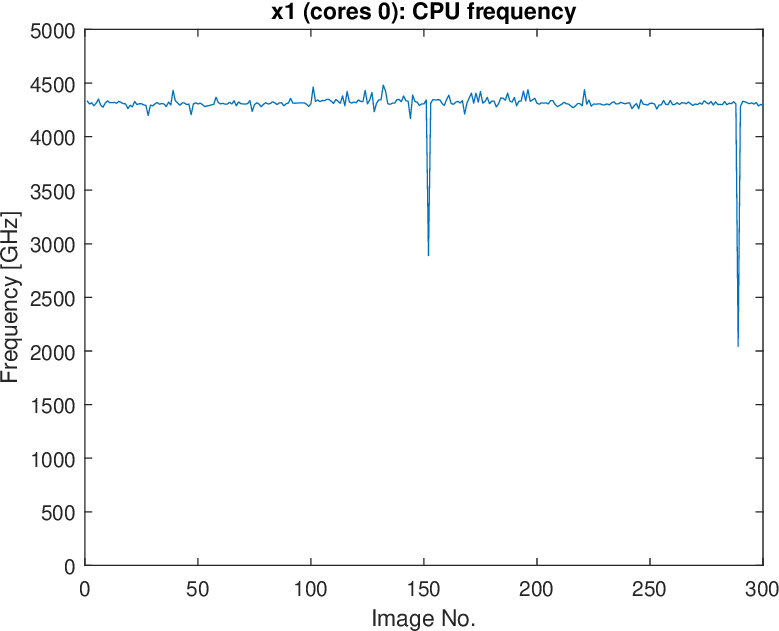,height=\plotsize}}
	\end{minipage}
	\caption{CPU characteristics for 1 instance of RAPiD executed on logical core \#0.}
	\label{fig:rapid0}
\end{figure*}

\begin{figure*}[!htb]
	\begin{minipage}[t]{0.48\linewidth}
		\centerline{\epsfig{figure=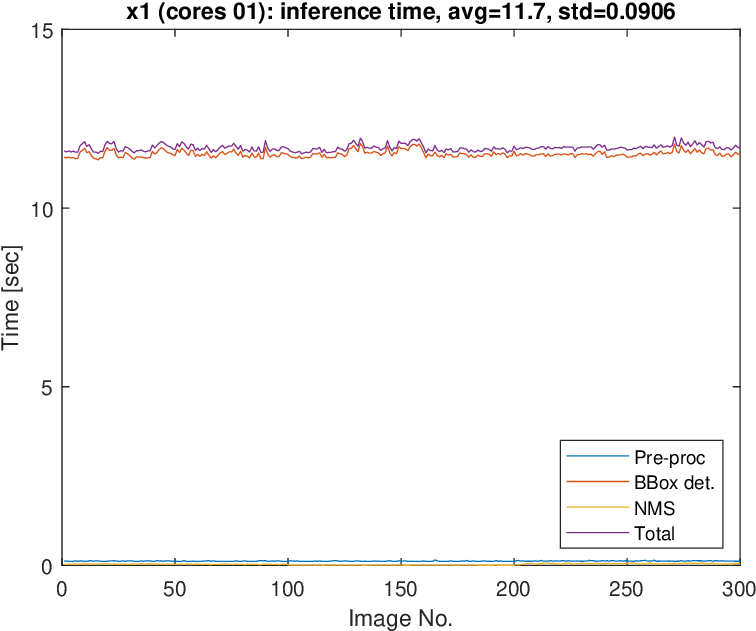,height=\plotsize}}
	\end{minipage}
	\hfill
	\begin{minipage}[t]{0.48\linewidth}
		\centerline{\epsfig{figure=plots/data_bboxes.eps,height=\plotsize}}
	\end{minipage}
	\bigskip\bigskip
	
	\begin{minipage}[t]{0.48\linewidth}
		\centerline{\epsfig{figure=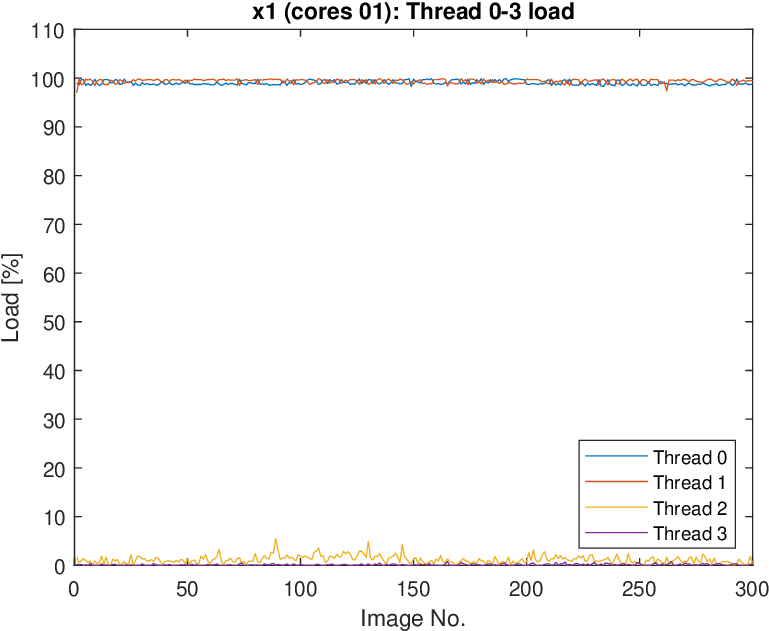,height=\plotsize}}
	\end{minipage}
	\hfill
	\begin{minipage}[t]{0.48\linewidth}
		\centerline{\epsfig{figure=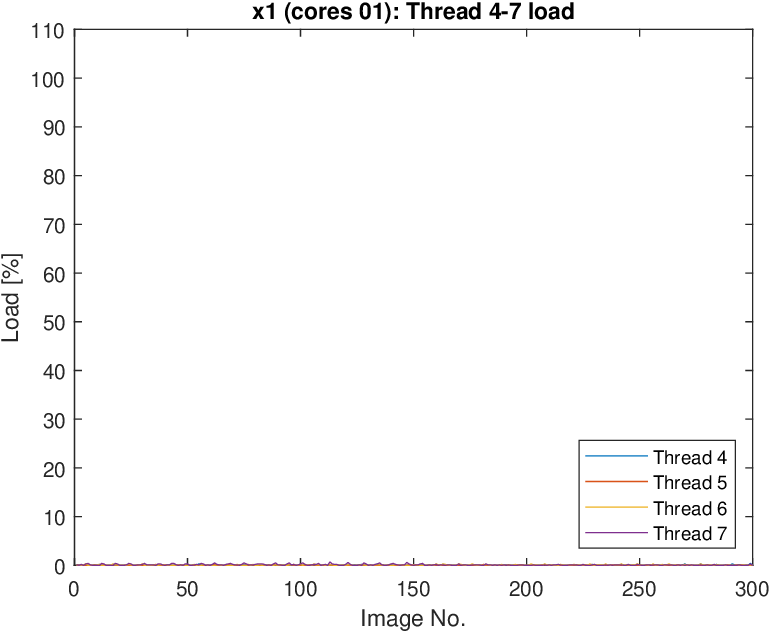,height=\plotsize}}
	\end{minipage}
	\bigskip\bigskip
	
	\begin{minipage}[t]{0.48\linewidth}
		\centerline{\epsfig{figure=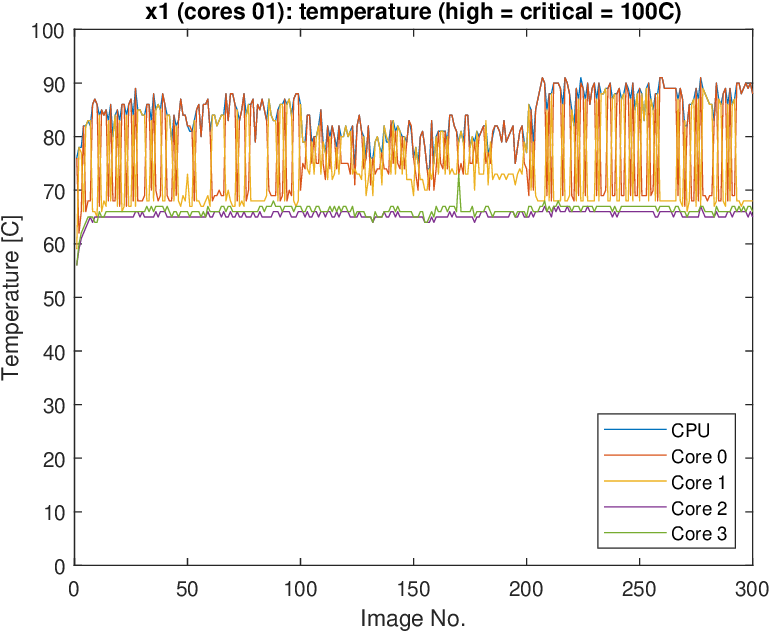,height=\plotsize}}
	\end{minipage}
	\hfill
	\begin{minipage}[t]{0.48\linewidth}
		\centerline{\epsfig{figure=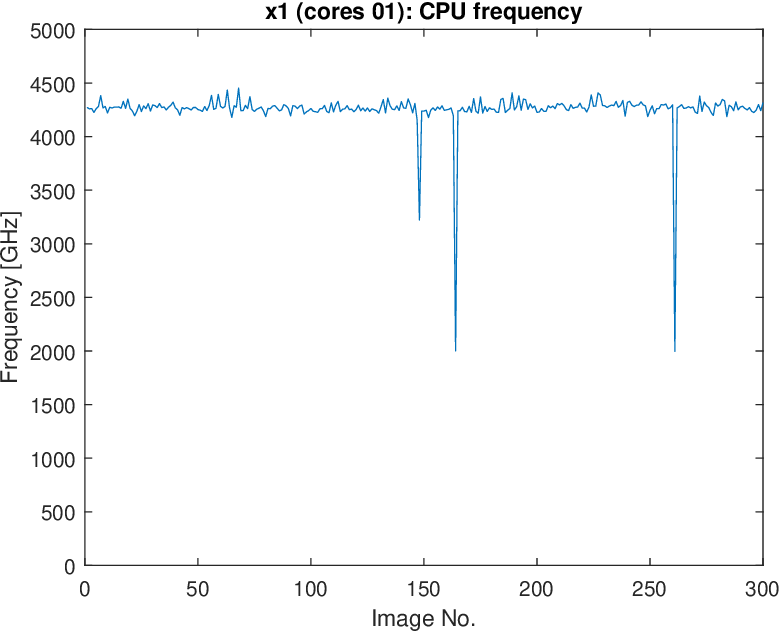,height=\plotsize}}
	\end{minipage}
	\caption{CPU characteristics for 1 instance of RAPiD executed on logical cores \#0-1.}
	\label{fig:rapid01}
\end{figure*}

\begin{figure*}[!htb]
	\begin{minipage}[t]{0.48\linewidth}
		\centerline{\epsfig{figure=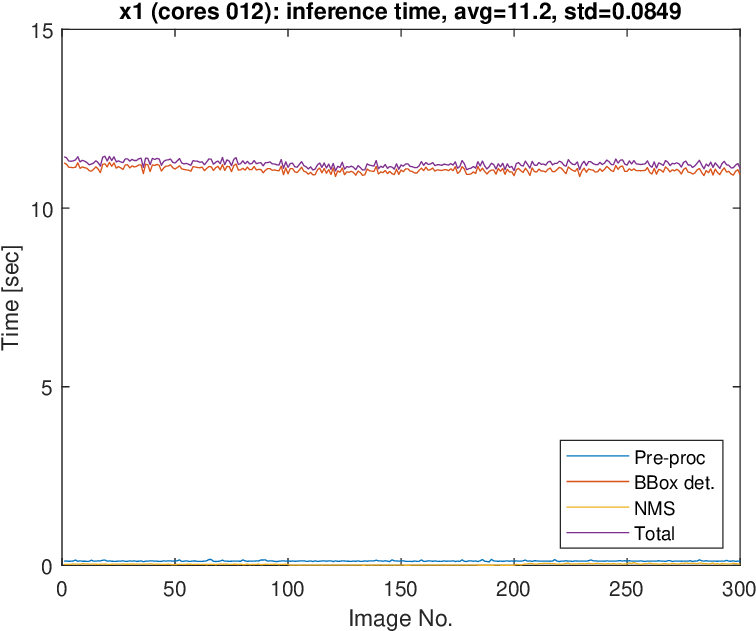,height=\plotsize}}
	\end{minipage}
	\hfill
	\begin{minipage}[t]{0.48\linewidth}
		\centerline{\epsfig{figure=plots/data_bboxes.eps,height=\plotsize}}
	\end{minipage}
	\bigskip\bigskip
	
	\begin{minipage}[t]{0.48\linewidth}
		\centerline{\epsfig{figure=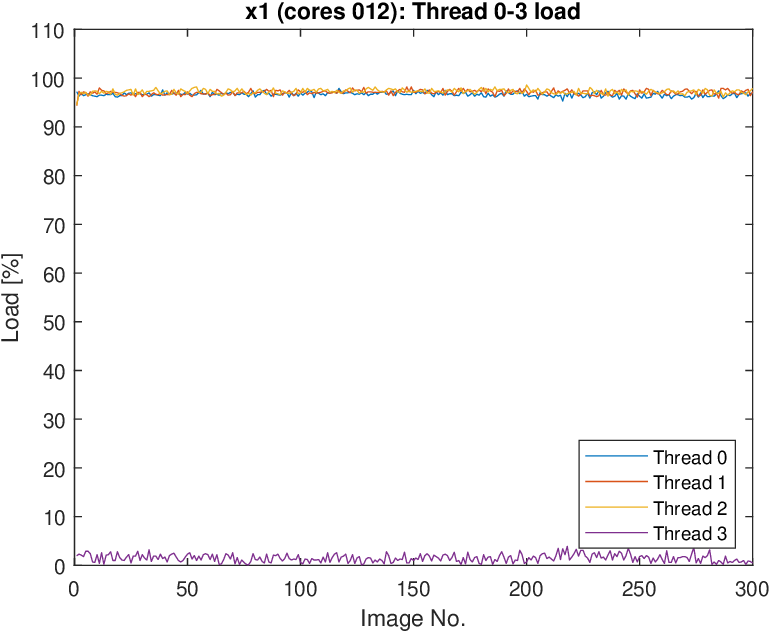,height=\plotsize}}
	\end{minipage}
	\hfill
	\begin{minipage}[t]{0.48\linewidth}
		\centerline{\epsfig{figure=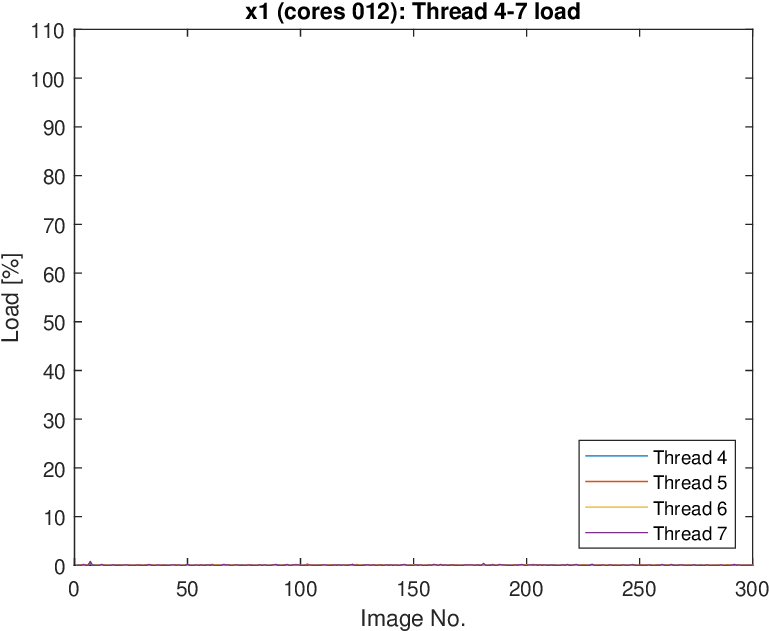,height=\plotsize}}
	\end{minipage}
	\bigskip\bigskip
	
	\begin{minipage}[t]{0.48\linewidth}
		\centerline{\epsfig{figure=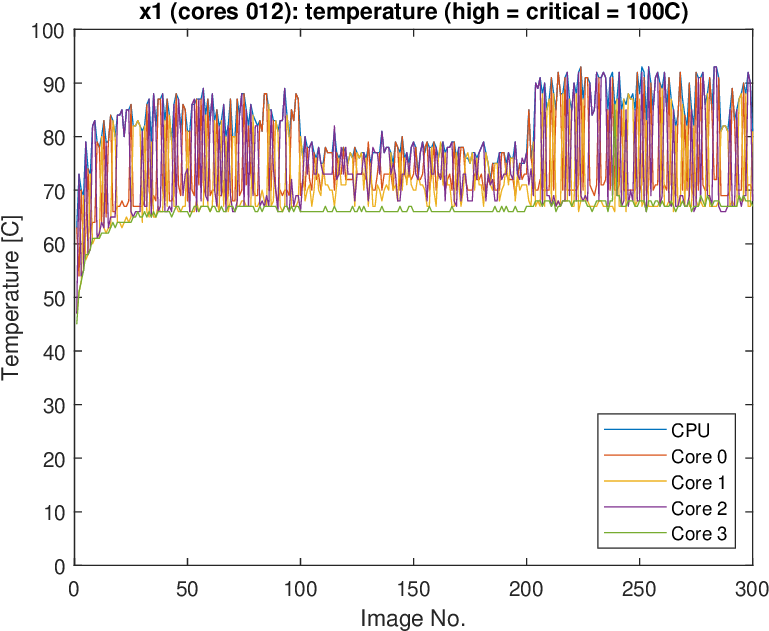,height=\plotsize}}
	\end{minipage}
	\hfill
	\begin{minipage}[t]{0.48\linewidth}
		\centerline{\epsfig{figure=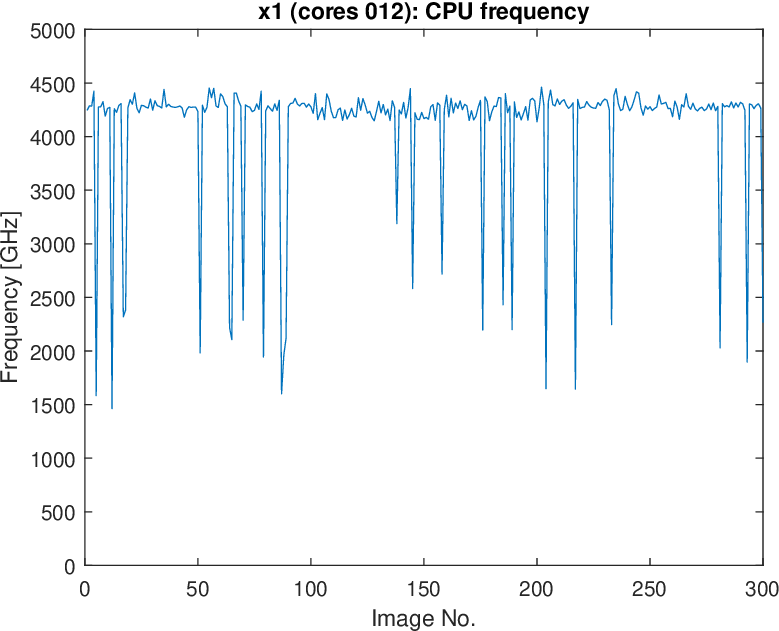,height=\plotsize}}
	\end{minipage}
	\caption{CPU characteristics for 1 instance of RAPiD executed on logical cores \#0-2.}
	\label{fig:rapid012}
\end{figure*}

\begin{figure*}[!htb]
	\begin{minipage}[t]{0.48\linewidth}
		\centerline{\epsfig{figure=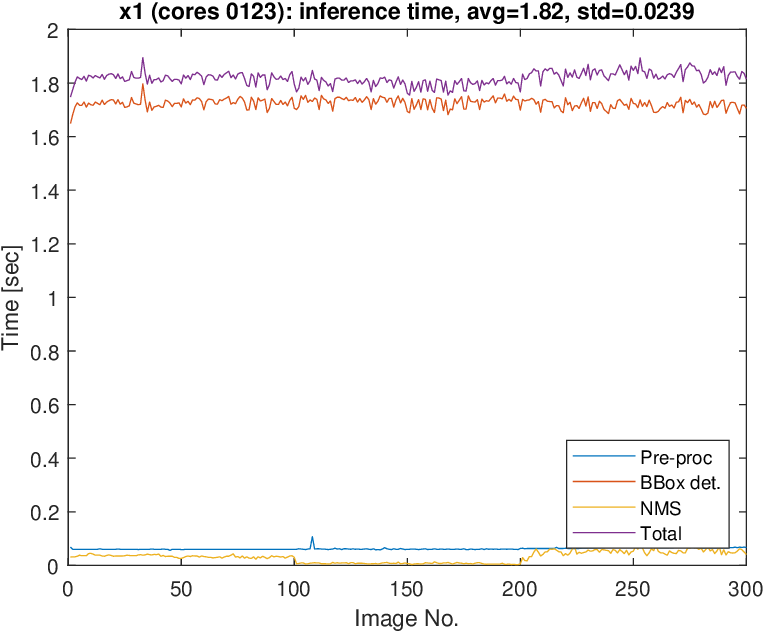,height=\plotsize}}
	\end{minipage}
	\hfill
	\begin{minipage}[t]{0.48\linewidth}
		\centerline{\epsfig{figure=plots/data_bboxes.eps,height=\plotsize}}
	\end{minipage}
	\bigskip\bigskip
	
	\begin{minipage}[t]{0.48\linewidth}
		\centerline{\epsfig{figure=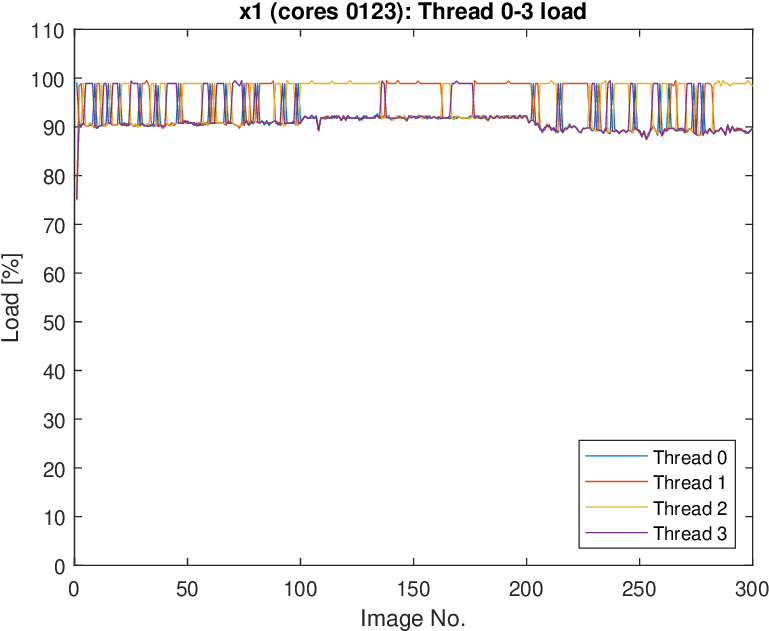,height=\plotsize}}
	\end{minipage}
	\hfill
	\begin{minipage}[t]{0.48\linewidth}
		\centerline{\epsfig{figure=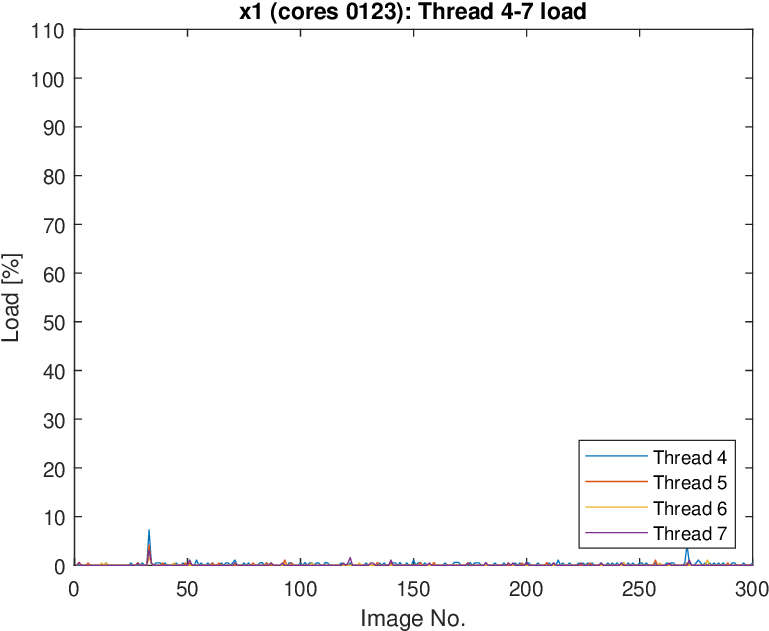,height=\plotsize}}
	\end{minipage}
	\bigskip\bigskip
	
	\begin{minipage}[t]{0.48\linewidth}
		\centerline{\epsfig{figure=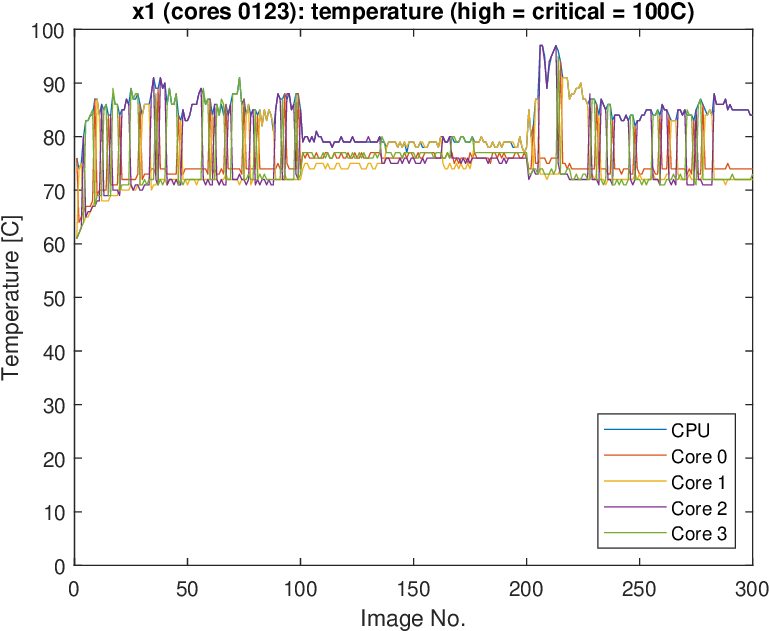,height=\plotsize}}
	\end{minipage}
	\hfill
	\begin{minipage}[t]{0.48\linewidth}
		\centerline{\epsfig{figure=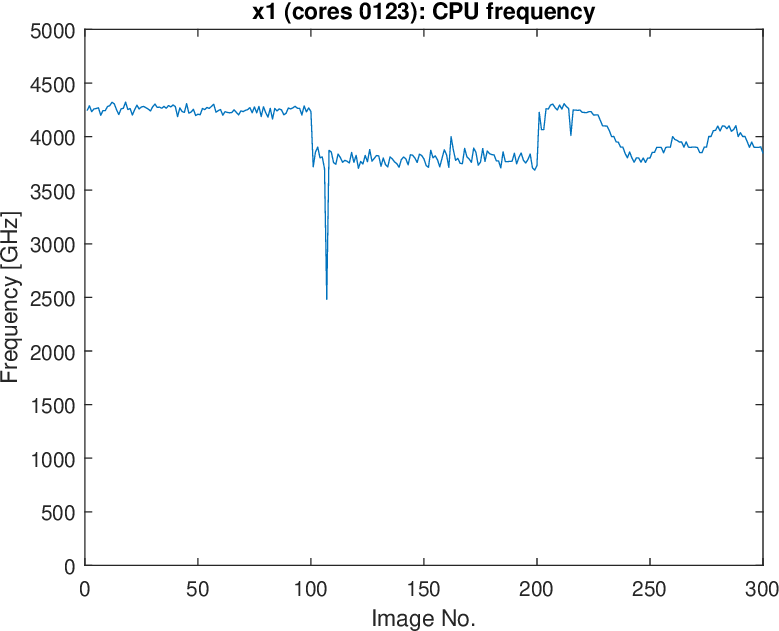,height=\plotsize}}
	\end{minipage}
	\caption{CPU characteristics for 1 instance of RAPiD executed on logical cores \#0-3.}
	\label{fig:rapid0123}
\end{figure*}

\begin{figure*}[!htb]
	\begin{minipage}[t]{0.48\linewidth}
		\centerline{\epsfig{figure=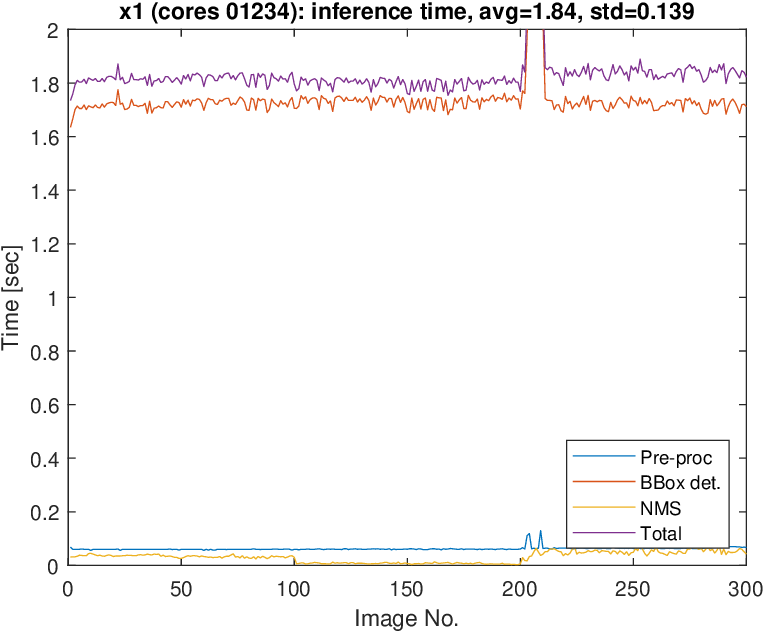,height=\plotsize}}
	\end{minipage}
	\hfill
	\begin{minipage}[t]{0.48\linewidth}
		\centerline{\epsfig{figure=plots/data_bboxes.eps,height=\plotsize}}
	\end{minipage}
	\bigskip\bigskip
	
	\begin{minipage}[t]{0.48\linewidth}
		\centerline{\epsfig{figure=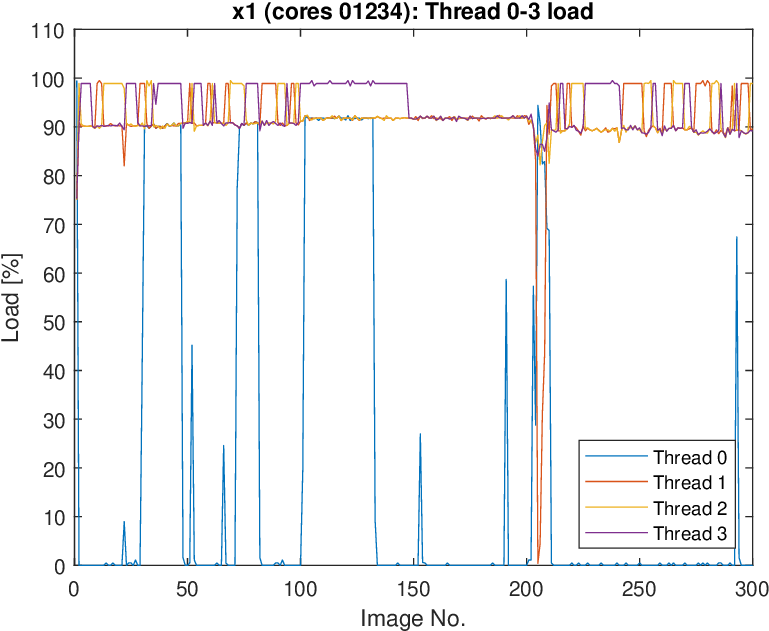,height=\plotsize}}
	\end{minipage}
	\hfill
	\begin{minipage}[t]{0.48\linewidth}
		\centerline{\epsfig{figure=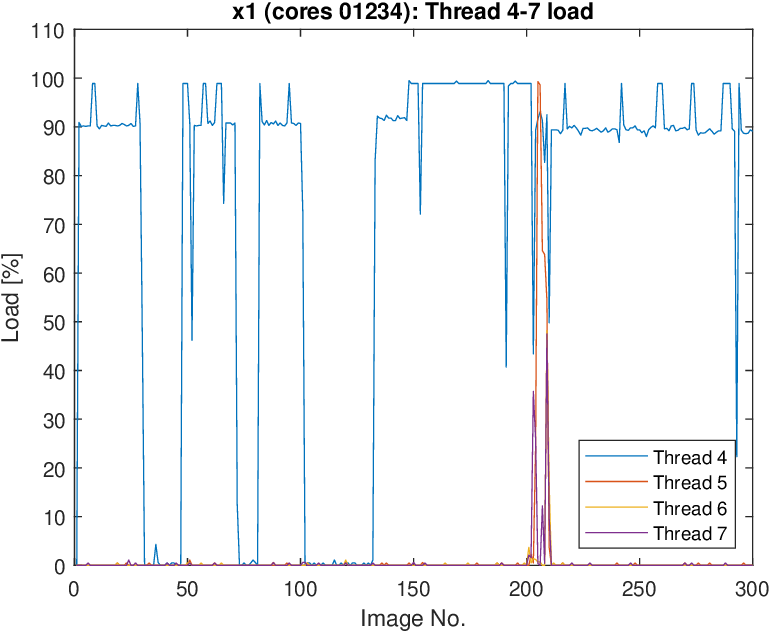,height=\plotsize}}
	\end{minipage}
	\bigskip\bigskip
	
	\begin{minipage}[t]{0.48\linewidth}
		\centerline{\epsfig{figure=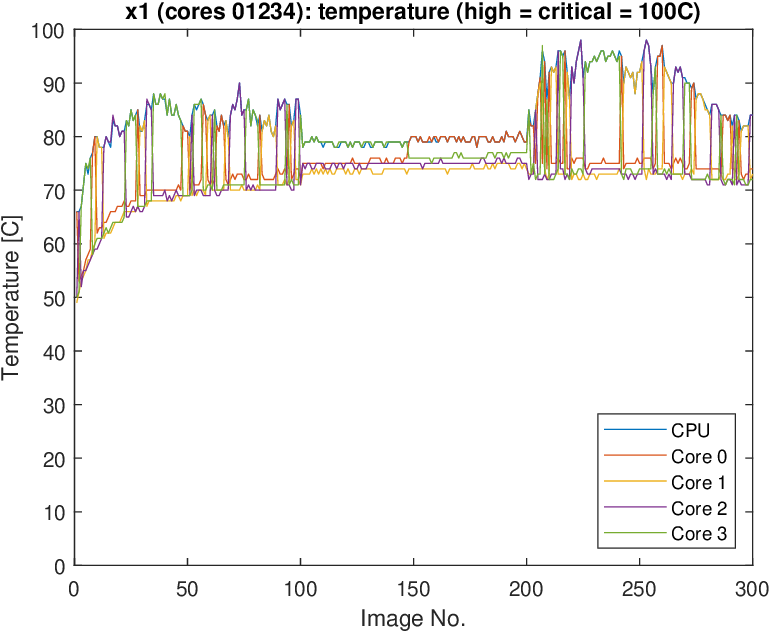,height=\plotsize}}
	\end{minipage}
	\hfill
	\begin{minipage}[t]{0.48\linewidth}
		\centerline{\epsfig{figure=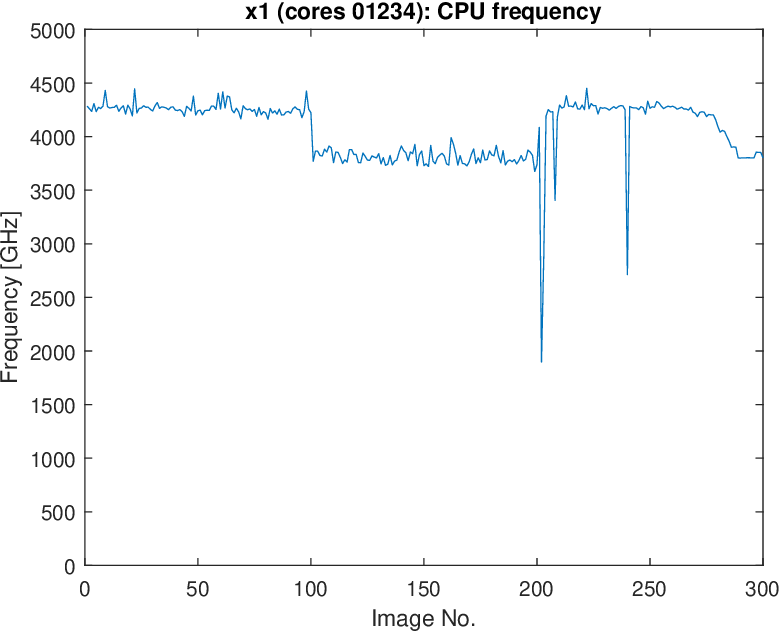,height=\plotsize}}
	\end{minipage}
	\caption{CPU characteristics for 1 instance of RAPiD executed on logical cores \#0-4.}
	\label{fig:rapid01234}
\end{figure*}

\begin{figure*}[!htb]
	\begin{minipage}[t]{0.48\linewidth}
		\centerline{\epsfig{figure=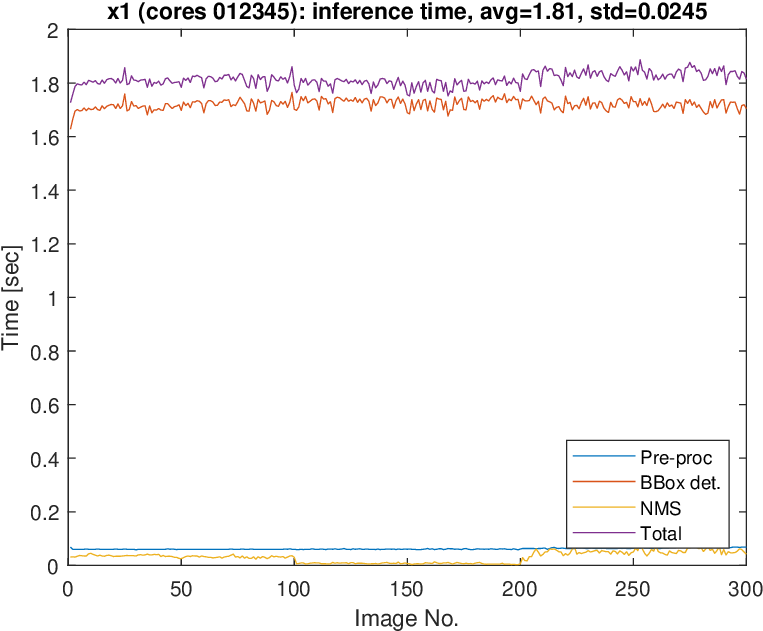,height=\plotsize}}
	\end{minipage}
	\hfill
	\begin{minipage}[t]{0.48\linewidth}
		\centerline{\epsfig{figure=plots/data_bboxes.eps,height=\plotsize}}
	\end{minipage}
	\bigskip\bigskip
	
	\begin{minipage}[t]{0.48\linewidth}
		\centerline{\epsfig{figure=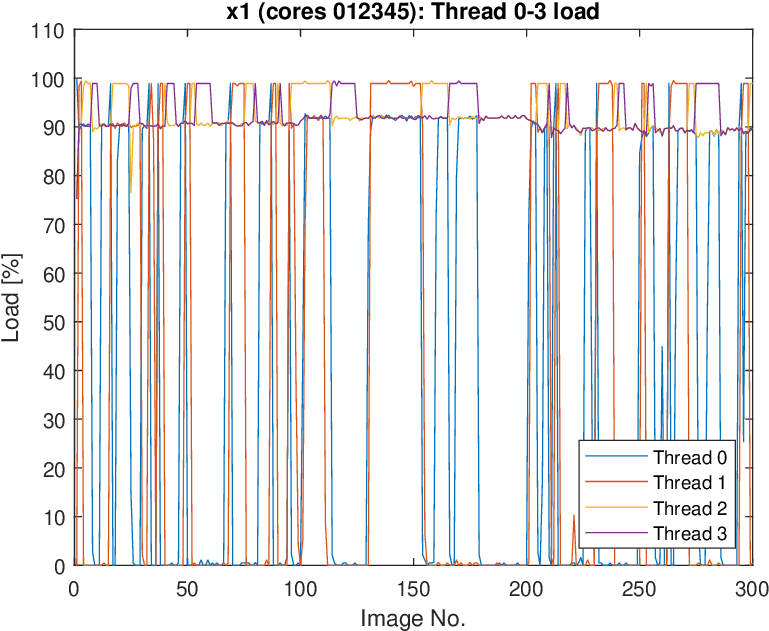,height=\plotsize}}
	\end{minipage}
	\hfill
	\begin{minipage}[t]{0.48\linewidth}
		\centerline{\epsfig{figure=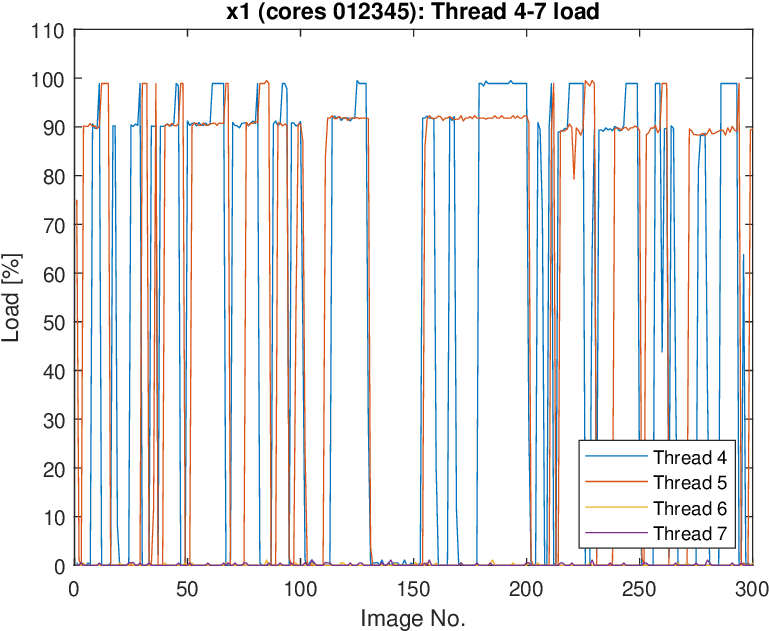,height=\plotsize}}
	\end{minipage}
	\bigskip\bigskip
	
	\begin{minipage}[t]{0.48\linewidth}
		\centerline{\epsfig{figure=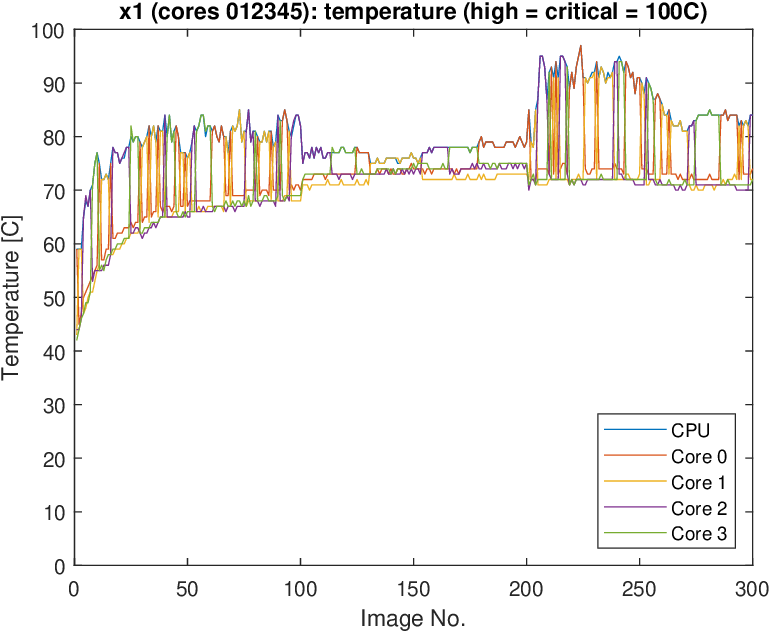,height=\plotsize}}
	\end{minipage}
	\hfill
	\begin{minipage}[t]{0.48\linewidth}
		\centerline{\epsfig{figure=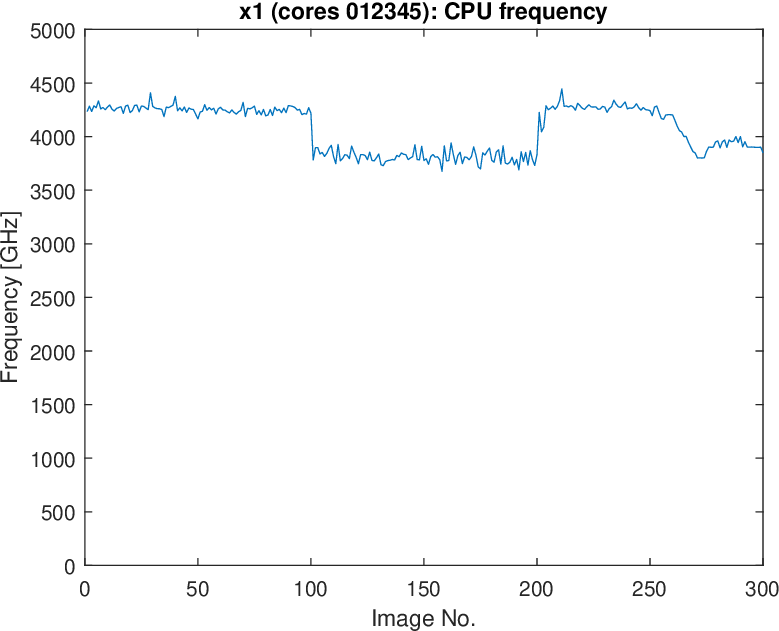,height=\plotsize}}
	\end{minipage}
	\caption{CPU characteristics for 1 instance of RAPiD executed on logical cores \#0-5.}
	\label{fig:rapid012345}
\end{figure*}

\begin{figure*}[!htb]
	\begin{minipage}[t]{0.48\linewidth}
		\centerline{\epsfig{figure=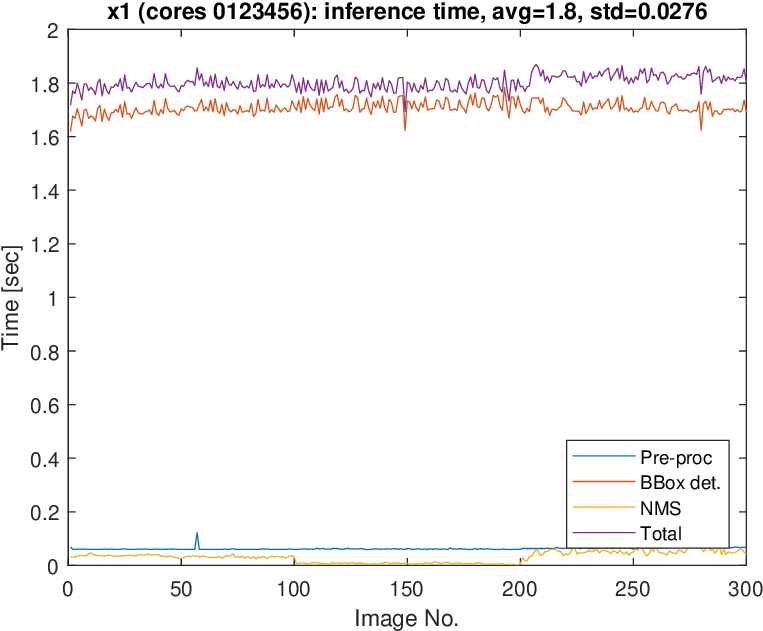,height=\plotsize}}
	\end{minipage}
	\hfill
	\begin{minipage}[t]{0.48\linewidth}
		\centerline{\epsfig{figure=plots/data_bboxes.eps,height=\plotsize}}
	\end{minipage}
	\bigskip\bigskip
	
	\begin{minipage}[t]{0.48\linewidth}
		\centerline{\epsfig{figure=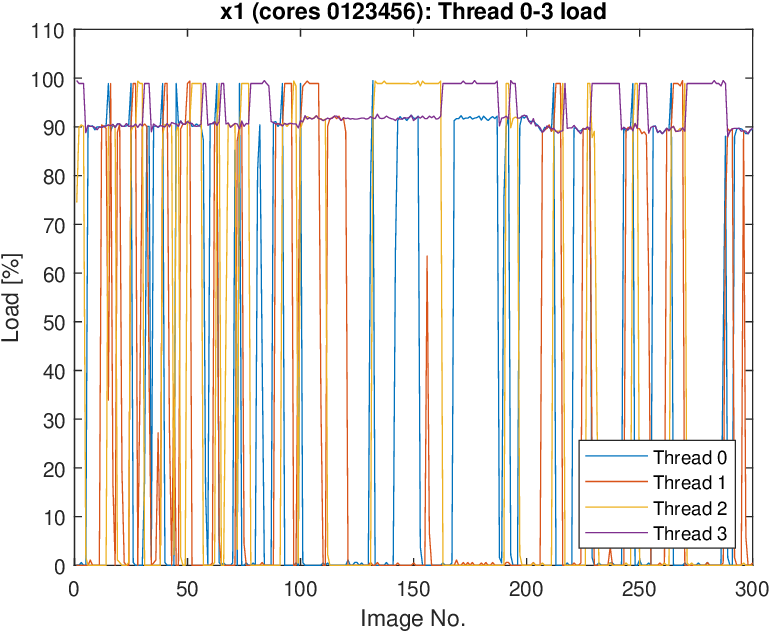,height=\plotsize}}
	\end{minipage}
	\hfill
	\begin{minipage}[t]{0.48\linewidth}
		\centerline{\epsfig{figure=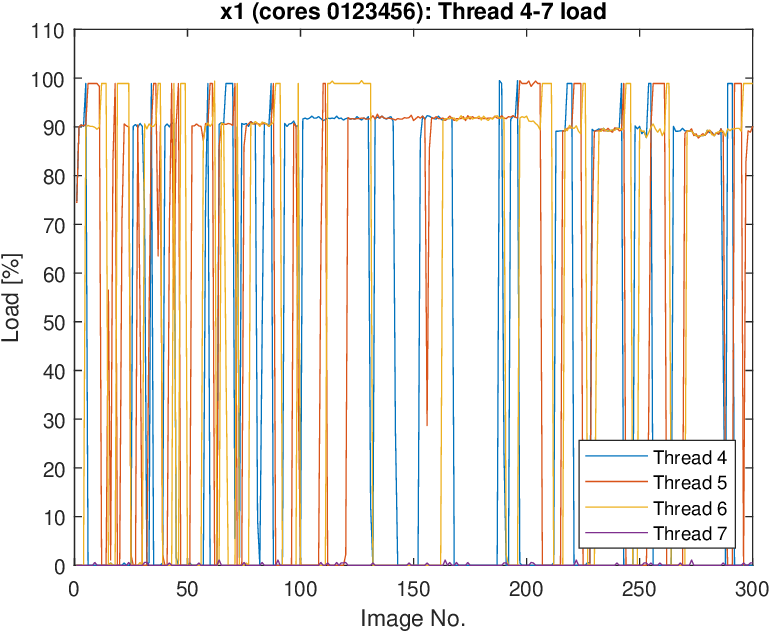,height=\plotsize}}
	\end{minipage}
	\bigskip\bigskip
	
	\begin{minipage}[t]{0.48\linewidth}
		\centerline{\epsfig{figure=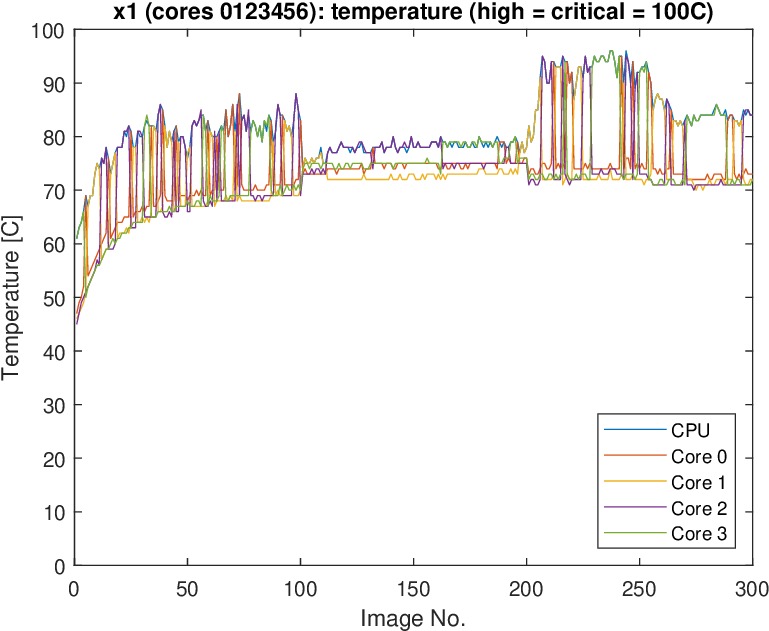,height=\plotsize}}
	\end{minipage}
	\hfill
	\begin{minipage}[t]{0.48\linewidth}
		\centerline{\epsfig{figure=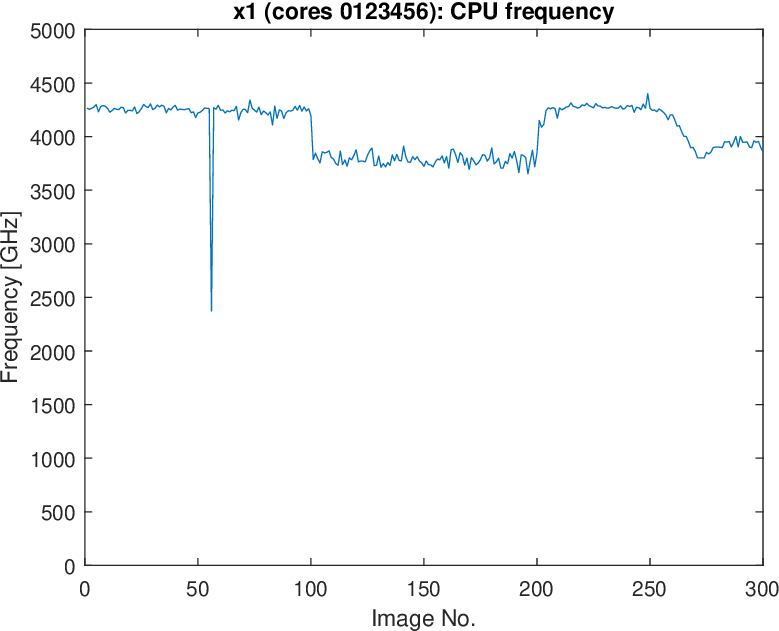,height=\plotsize}}
	\end{minipage}
	\caption{CPU characteristics for 1 instance of RAPiD executed on logical cores \#0-6.}
	\label{fig:rapid0123456}
\end{figure*}

\end{document}